\documentclass[12pt]{article}
\usepackage{latexsym,graphicx,natbib,aas_macros,placeins,amsmath,amssymb,setspace}

\newcommand{\beq}{\begin{equation}}
\newcommand{\eeq}{\end{equation}}

\begin{document}
\bibliographystyle{icarus}

\title{A deeper  look at the colors of the Saturnian irregular satellites}

\medskip

\author{Tommy~Grav \\ Harvard-Smithsonian Center for Astrophysics, \\ MS51, 60 Garden St., Cambridge, MA02138 \\ 
and \\ Instiute for Astronomy, University of Hawaii, \\ 2680 Woodlawn Dr., Honolulu, HI86822 \\
{\it tgrav@cfa.harvard.edu} \\ \\
James~Bauer \\ Jet Propulsion Laboratory, California Institute of Technology \\ 4800 Oak Grove Dr., MS183-501, Pasadena, CA91109 \\{\it bauer@scn.jpl.nasa.gov}}
\maketitle

\vspace{5cm}
{Manuscript: 36 pages, with 11 figures and 5 tables.}

\newpage
\noindent Proposed running head: Colors of Saturnian irregular satellites \\

\vspace{2cm}
Corresponding author:

\vspace{0.5cm}
\indent\indent Tommy Grav \\
\indent\indent MS51, 60 Garden St. \\
\indent\indent Cambridge, MA02138 \\
\indent\indent USA \\
\indent\indent Phone: (617) 384-7689 \\
\indent\indent Fax: (617) 495-7093 \\
\indent\indent Email: tgrav@cfa.harvard.edu


\newpage
\begin{abstract}
We have performed broadband color photometry of the twelve brightest irregular satellites of Saturn with the goal of understanding their surface composition, as well as their physical relationship. We find that the satellites have a wide variety of different surface colors, from the negative spectral slopes of the two retrograde satellites S~IX~Phoebe ($S'=-2.5\pm0.4$) and S~XXV~Mundilfari ($S'=-5.0\pm1.9$) to the fairly red slope of S~XXII~Ijiraq ($S'=19.5\pm0.9$). We further find that there exist a correlation between dynamical families and spectral slope, with the prograde clusters, the Gallic and Inuit, showing tight clustering in colors among most of their members. The retrograde objects are dynamically and physically more dispersed, but some internal structure is apparent. 

\end{abstract}

\vspace{2cm}
Keywords: Irregular satellites; Photometry, Satellites, Surfaces; Saturn, Satellites.
\newpage
\section{Introduction}
 
The satellites of Saturn can be divided into two distinct groups, the regular and irregular, based on their orbital characteristics. The regular satellites are believed to have been created in a scenario similar to the Solar System, accreting out of a circumplanetary disk of gas and dust. They are thus on prograde, nearly circular orbits close to the equatorial plane of their host planet. The irregular satellites are thought to be captured heliocentric objects, but their origin or capturing processes are not well understood. 

The first irregular satellite of Saturn was S~IX~Phoebe, discovered by W. Pickering in 1898. It would not be until a full century later that the next irregular satellite of Saturn was announced, when \citet{Gladman.2001a} reported their discovery of 12 new irregular satellites, ranging in size from 5 to 35 km in diameter. Subsequent surveys by \citet{IAUC.8116,IAUC.8727} and \citet{IAUC.8523} have raised the number of known Saturnian irregular satellites to a total of 35. 

The irregular satellites of Saturn were recognized to be clustering in orbital elements by \citet{Gladman.2001a}. They suggested that the clustering was due to the breakup of several larger progenitors, similar to the creation of the asteroidal families. The Saturnian irregular satellites seem to make up two clear prograde families, named after Inuit and Gallic gods respectively, and a much more dispersed retrograde group, named after the Norse ice giants. It is not clear whether theses retrogrades are fragments of one or more captured progenitors, although the great dispersion of the 18 known retrograde satellites makes the idea of several progenitors more likely. To test the hypothesis that the dynamical families are indeed fragments of larger progenitors \citet{Grav.2003b} presented $BVRI$ photometry of 8 saturnian irregular satellites. The prograde families were indeed found to have homogeneous colors, but only 3 of the objects had 1-$\sigma$ uncertainty less than $0.05$ across all color, making it hard to strongly argue in favor of the hypothesis proposed by \citet{Gladman.2001a}. 

In order to place constraints on the many capturing mechanisms that have been proposed to cause the creation of the irregular satellite population, it is necessary to understand the number of events needed to create the currently observed dynamical structure. Irregular satellites are generally considered to have been captured by their host planets from heliocentric orbits \cite{Gladman.2001a}. Due to the time reversibility of the pure gravitational three-body capture (sun-planet-satellite) the object is only temporarily in orbit around the planet unless some dissipative mechanism exist to make the capture permanent. At least five different processes have been proposed: 1) capture due to collisions between a temporarily captured object and an existing satellite, or between two temporarily captured objects \citep{Colombo.1971,Gladman.2001a}; 2) capture due to a mass increase in the host planet \citep{Heppenheimer.1977}; 3) capture due to gas drag from a gas envelope or gas disk around the planet \citep{Pollack.1979,Cuk.2004a}; 4) capture through three-body interaction with other temporarily captured objects \citep{Jewitt.2005b}; 5) capture during planetary resonance passage \citep{Cuk.2006a}. It is evident that a better understanding of the physical and dynamical structure of the irregular satellites, as well as determining the number of capturing events, is essential in constraining and potentially eliminating some of these proposed mechanisms. Note that it is still not clear whether one mechanism dominates the capture, or whether the irregular satellites observed today are the result of capture through several of the proposed mechanisms. 

The study of the physical characteristics of the irregular satellites is also an integral part in determining the origins of the progenitors. Understanding where the irregular satellites originate from and how they reached the phase space of the giant planets is essential in order to improve our understanding of the transfer of matter within the early Solar System. Several origins have been proposed for the irregular satellites. \citet{Cuk.2004a} proposed that the Himalia family originated among the Hilda asteroids in the outer main asteroid belt.  The recent flyby of S~IX~Phoebe by the Cassini spacecraft just 19 days before orbital insertion into the Saturnian system provided a wealth of new information. \citet{Johnson.2005a} argued that the measured density of Phoebe, at $\rho \sim 1630\pm45 \mbox{kgm}^{-3}$, indicates that Phoebe has its origin in the outer Solar System. Origins among the trojan population and local population of planetesimals at the later stages of giant planet formation has also been proposed. 

This paper presents optical photometry collected using the GMOS-N imager on the 8m Gemini North telescope and the LRIS-R\&B imager on the the 10m Keck I telescope in preparation for thermal observations of the irregular satellites with the MIPS instrument on the Spitzer Space Telescope. It also provides an errata to a measurement of Albiorix published in \citet{Grav.2003b}.

\section{Observations}

Optical photometry of 7 Saturnian irregular satellites in the $g'$, $r'$, $i'$ and $z'$ SDSS-filters were collected through queue observations on November 9th and 10th, 2004 and classical observing on April 13th and 15th, 2005, both using the Gemini Multi-Object Spectrograph North (GMOS-N) instrument at the 8-m Gemini Observatory on Mauna Kea, Hawaii. GMOS-N has a $\sim 5$ by $\sim 5$ arcmin field of view, with $0.073$ arcsec per pixel. All observations with this instrument were done in binned mode ($0.146$ arcsec per pixel), which with seeing in the range $0.75-1$ arcsec gave excellent sampling of the point spread function of our targets.  Optical photometry of 8 Saturnian satellites were also collected in the Johnson-Kron-Cousins BVRI filters on January 6th, 2005, using the Low Resolution Imaging Spectrometer (LRIS) on the 10m Keck I telescope, also located on Mauna Kea, Hawaii. LRIS has two detectors, the red and blue side of the instrument, allowing observations in two filters simultaneous. The red side has a field of view of $\sim7$ by $\sim 7$ arcmin, with a plate scale of $0.21$ arcsec per pixel. The blue side has field of view of $\sim 9$ by $\sim 9$ arcmin, with a plate scale of $0.135$ arcsec per pixels. In addition we report the re-reduction of one observation from the \citet{Grav.2003b} that was found to be erroneous. The observational circumstances are given in Table \ref{tabphysical}. 

The data were collected by observing each object in a RBBRVVRIIR sequence (or the equivalent for the SDSS filter set), to make sure that rotational light curve effects did not influence the computed colors. All images were taken with sidereal tracking and the exposure time was kept short enough to avoid trailing of the satellites. Images in each of the filters were then combined to increased the signal-to-noise ratio of the science targets. 

The collected data was reduced using aperture corrected photometry with the IRAF/DAOPHOT package \citep{IRAF, DAOPHOT}. Science target aperture radii were selected at $\sim 1.3\times\mbox{fwhm}$ and aperture correction values were determined using outer aperture radii of $3-4\times\mbox{fwhm}$ on $\sim8-12$ field stars. Background flux were calculated using an aperture radius from $\sim4$ to $\sim8$ times the full-width-half-max (fwhm) of the fields point spread function (psf). Zero point magnitudes, airmass and color corrections were determined using a set of \citet{Landolt.stand} standard stars taken on the same nights as the science observations. In order to compare the resulting photometry the SDSS filtered observations were converted to the BVRI filter system. Transformation equations between the SDSS photometric system ($g'r'i'z'$) and the standard Johnson-Kron-Cousins system \citep{Johnson.1953a,Cousins.1978a} are given in \citet{Fukugita.1996a} and \citet{Jordi.2006a}.   

\section{Results}
\label{sec:results}

With this survey we have performed accurate color broadband photometry of the 12 most luminous Saturnian irregular satellites. We have observed all but one of the prograde satellites (S/2004~S11 in the Gallic family was not observed) and 1/5 of the known retrograde satellites of Saturn. This survey thus constitutes the most comprehensive study of color photometry of the irregular satellites to date. The results are given in Table \ref{tabresults} and show that the colors and spectral slopes of the Saturnian irregular satellites are consistent with C-, D- and P-type asteroids. Solar colors were adopted as $B-V=0.67$, $V-R=0.36$ and $V-I=0.71$ \citep{Hartmann.1982a,Hartmann.1990a,Hardorp.1980a,Jewitt.1998a}.

The colors were converted to reflectance and the resulting broadband spectra are given in Fig. \ref{fig.spectra_gallic}, \ref{fig.spectra_inuit} and \ref{fig.spectra_nordic} for the Gallic, Inuit and retrograde clusters, respectively. The spectral slope, $S_1' [\% \mbox{per} 100 \mbox{nm}]$, of the broadband colors are given by 
\begin{equation}
  {S'}_1 = \frac{2 (10^{0.4(C - C_{sun})} - 1)}{\Delta \lambda (10^{0.4(C - C_{sun})} + 1)} 
\end{equation}
where $C$ and $C_{sun}$ are the colors of the object and sun, respectively. $\Delta \lambda$ is the wavelength difference between the two filters that make up the color \citep{Luu.1996a}. The value given for ${S'}_{1}$ in Table \ref{tabslopes} are found using the $B-I$ colors of the objects and the sun.  Since some of the satellites have apparent non-linear features, we also fitted the four reflectances to a linear slope ${S'}_2$ using a linear regression method. Note that for the non-linear slopes of S~XXIII~Suttungr and S~XXX~Thrymr the fit was done excluding the $V$ filter reflectance. Fitting the ${S'}_2$ gave similar values to the ${S'}_1$ calculation. However, ${S'}_2$ is generally a better way to derive an average slope solution when the normalized reflectance's deviate significantly from a linear solution, as it uses all four points (or in some cases three points) rather than the two used in calculating ${S'}_1$. 

To capture the non-linearity we also fit linear slopes for filter subsets, the $BVR$ filters, ${S'}_{2(B-R)}$, and for the $RI$ filters, ${S'}_{2(R-I)}$.  All four values for the spectral slope is given in Table \ref{tabslopes}. Using the spectral slope to taxonomy class values given in \citet{Dahlgren.1995a}, we define taxonomy classes for our objects (again see Table \ref{tabslopes}). It should be noted that the spectral slopes used in \citet{Dahlgren.1995a} are based on the spectral slope in the range $400-740$nm, while our slopes span the range $430-900$nm. Since most of the slopes are near linear, we can still safely use the literature values for comparison. The results are plotted in Fig. \ref{fig.spectralslope} and show that the Saturnian irregular satellites contain a more or less equal fraction of C-, P- and D-like objects. One object, Ijiraq, has a spectrum that is marginally redder than the D-type objects and we define this as an "medium red"-type object (MR-type; $15\% < {S'}_2 < 25\%$). No objects appears to have the ultra-red matter (UR-type, ${S'}_2 > 25\%$) found in the trans-neptunian population \citep[e.q. ][]{Jewitt.2002a}. 

We also look for evidence for the weak absorption feature near $0.7 \mu$m that has been attributed to an $\mbox{Fe}^{2+} \rightarrow \mbox{Fe}^{3+}$. This feature has been shown to appear in spatially resolved spectra \citep{Jarvis.2000a} and in $\sim50\%$ of the main-belt C-type asteroids through both reflectance spectra and photometry \citep{Vilas.1994a,Bus.2002a,Carvano.2003a}. Using data from 27 low-albedo asteroids, \citet{Vilas.1994a} showed that there were a $\sim85\%$ correlation between the $0.7 \mu$m and the $3.0 \mu$m spectral feature due to water hydration. The weighted mean values of the $0.7\mu$m test for all available observations are given in Table \ref{tabphase}. 

The observed magnitudes were reduced to absolute magnitudes using the H-G formulation
\begin{equation}
     H(1,1,0) = m_V - 5 \log (r \Delta) + 2.5 \log ( (1 - G)\Phi_1(\alpha) + G \Phi_2(\alpha))
\end{equation}
where the first correction term is the distance modulus, with $r$ and $\Delta$ representing the object's distance in AU from the Sun and observer, respectively, and the second correction term is the phase modulus \citep[for a more thorough description see][]{Bowell.1989}. Using the available observations (both from this paper and literature in general) of the saturnian irregular satellites we fit $H(1,1,0)$ and $G$ using a Levenberg-Marquardt least square method. These values are given in Table \ref{tabphase} and the fits are plotted in Fig. \ref{fig.phasegallic} to \ref{fig.phasenordic}. Note that use of photometry reported to the Minor Planet Center (MPC)is generally discouraged for any serious application as it is generally unreliable and no information on the actual filters used is available. However, the MPC photometry used in this paper to derive phase curve parameters are well known to the authors, as they are all from survey and follow-up programs in which the authors are members \citep{Gladman.2001a}. These observations have a signal to noise ratio of $\sim 4$ in most cases, thus we assume that the error in the MPC collected photometry is $0.25$ magnitudes. It should be noted that several of the less observed objects still have very poor solutions and additional photometric observations are still badly needed to better constrain and utilize the  phase curve as a marker for relationships between the irregular satellites of Saturn. However, using the fitted value for the absolute magnitude and assuming an albedo of $p_R = 0.08$ \citep[the average albedo of Phoebe; ][]{Simonelli.1999a} we calculate the approximate diameter of the saturnian irregular satellites observed in this paper (see Table \ref{tabphase}). It should also be noted that without knowledge about the rotational light curve caution should be taken in reading to much into the phase curve parameter. For example, data collected by the authors do show that Ymir has a large rotational light curve amplitude ($\sim 0.3$ peak-to-peak magnitudes) and, while the phase curve fit does not show much scatter around the fit, care is needed when interpreting the result. It should be noted that arguments about the inadequacy of the H-G formulation in detecting the opposition surge for very small phase angles has been made, and several other groups prefer the use of a combination of a linear and exponential function with four free parameters \citep{Belskaya.2006a,Kaasalainen.2003a}. With additional observations such models could be used to better model the apparent strong opposition surge of Paaliaq and Ymir (see Figure \ref{fig.phaseinuit} and \ref{fig.phasenordic}, but with the sparse photometry currently available the 2 free parameter of the H-G formulation remains an adequate method for determination of the phase light curve. 

In order to study the collisional events as a possible mechanism behind the creation of the apparent dynamical family structure, we used the Gauss  equations to calculate the dispersion velocity needed to moved the orbits of the fragments from that of the progenitor \citep{Nesvorny.2003,Nesvorny.2004,Grav.2003b}. We further use the same method as \citet{Nesvorny.2004} to compute the size of the impactor needed to create such a dispersion velocity assuming that $10\%$ of the impact energy is deposited into the fragments as kinetic energy. Using the derived sizes from Table \ref{tabphase} and an assumed density of $\rho = 1.6 \mbox{ g/cm}^3$ \citep{Johnson.2005a} for all bodies we can determine a likely scenario for the collision and fragmentation. Since the collision rates among the satellites or with cometary objects are too low to explain the apparent family structures \citep{Nesvorny.2004}, we use remnant planetesimals from early epochs of the Solar System in an attempt to determine the disk mass needed to create the families. We use the same three size distributions as \citet{Nesvorny.2004}, and note that the size distribution based on studies of the in situ formation of the Kuiper belt \citep{Stern.1995,Kenyon.1998,Kenyon.2002} is an unlikely analog to a disk of remnant planetesimals at Saturn. \citet{Nesvorny.2004} showed that this distribution (called C in their work) would catastrophically disrupt even the smallest of the known Saturnian irregular satellites, unless the disk have a mass $M_{\mbox{disk}} \lesssim 10 M_\oplus$. 

\subsection{The Gallic Family}
\label{sec:gallic} 

This dynamical group is centered around $37^\circ$ inclination and currently four members are known (see Table \ref{taborbele}). Their semi-major axes are also tightly clustered, $a_{\mbox{\tiny avg}} = 207-302$ saturn radii ($R_S$), compared to the large spread among all the known Saturnian satellites, $a_{\mbox{\tiny avg}} = 186-412 R_S$. The largest member, S~XXVI~Albiorix, is also the object with the lowest semi-major axis among the group, $a_{\mbox{\tiny avg}} = 271 R_S$. This provides evidence that any fragmentation must have happened after any substantial gas envelope dissipated. All the members have high average eccentricity, $e_{\mbox{\tiny avg}} = 0.46-0.53$, which together with the tight clustering in inclination, strongly suggest that the four objects have a common origin. 

We observed three of the four known members of this family and find, contrary to our earlier work \citep{Grav.2003b}, that the family does not have homogenous colors (see Fig. \ref{fig.colors_gallic}). The new observations show that Albiorix has a significant steeper spectral slope, at ${S'}_2 = 12.5\pm0.4\%$, compared to S~XXI~Tarvos and S~XXVIII~Erriapo, with slopes of ${S'}_2 = 5.4\pm1.0\%$ and ${S'}_2 = 5.1\pm0.7\%$ respectively. This results does not  support the hypothesis that the three objects are fragments of a larger progenitor. However, a closer look at the individual observations show that Albiorix has a large span of different spectral slopes, from ${S'}_2 = 5.3\pm0.8\%$ in the observations on January 6th, 2005, to ${S'}_2 = 14.9\pm0.5\%$ in the observation on April 13th, 2005. It is thus evident  that the colors of Albiorix varies over its surface. Figure \ref{fig.spectra_gallic} shows the reflectance spectra of these three objects. It is easily seen that Tarvos and Erriapo have linear slopes to the $1\sigma$ confidence level, while Albiorix has a linear slope at a $3-\sigma$ (lower) confidence level in the January 2005 observations. The April 2005 observation, however, is significantly non-linear with a very steep slope from the B- to the R-filter (${S'}_{2(B-R)} = 19.0\pm0.7$) and a shallower R- to I-filter slope (${S'}_{2(R-I)} = 5.6\pm1.2$). The two new observations of Albiorix reported in this paper are displayed separately in Fig. \ref{fig.spectra_gallic} clearly showing their differences. The slope of the January 2005 observation of Albiorix is remarkable similar to the slopes of both Tarvos and Erriapo. We thus hypothesize that Tarvos and Erriapo are the largest fragments from an impact on Albiorix, leaving a less red crater. The phase curves of this family further supports the notion that the objects are indeed of similar origin (see Fig. \ref{fig.phasegallic}). Their phase curves are all fairly shallow, with a weighted mean phase slope of $G = 0.39$. The Gallic irregular satellites thus span the range for main belt asteroids \cite[$G \sim 0.1-0.5$;][]{Bowell.1989}.  Additional observations are needed to further constrain the phase curve fits for both Tarvos and Erriapo. 

The question is now whether a cratering event is possible for Gallic satellites. We find that a dispersion velocity of $\sim 90$ m/s needed to displace both the two smaller satellites to their current orbits from that of Albiorix. Such an impact could have been the result of $\sim 1.25$km planetesimal hitting with a collisional velocity of $4.79$ km/s. The impact would be just sub-catastrophic, excavating a crater with a radius of $\sim12$ km.  Such a collision has a $50\%$ probability of happening for a disk with $M_{\mbox{disk}} \gtrsim 50-100 M_\oplus$, depending on the size distribution of planetesimals. While this is a substantial disk, the many large craters on Phoebe is further evidence of such a massive disk \citep{Porco.2005}.

It is important to note that \citet{Cuk.2004b} found that the primordial dynamical distribution of this family of objects may have been significantly altered when the Great Inequality resonance ($\xi = 5\lambda_S - 2 \lambda_J - 2 \varpi - \varpi_S$) swept through the region during the migration of Jupiter and Saturn to their present configuration. This of course complicates our understanding of this family, and as mentioned by \citet{Cuk.2004b} more study of the impact of this resonance on the Gallic family is indeed needed. In particular, it would significantly help our understanding of the evolution of the irregular satellites to know whether this resonance favors the phase space currently inhabited by the Gallic family for capture.  If so, it could be that the Gallic cluster is populated by objects that during the sweeping of the Great Inequality resonance during planet migration was moved into the cluster from a larger region of the surrounding phase space. Such a study, coupled with the determination of any variations in color on the surface of Albiorix, would indicate whether these objects are indeed fragments from a cratering event on Albiorix, rather than the products of individual captures. 

\subsection{The Inuit Family}
\label{sec:inuit}

The Inuit cluster is centered around $\sim 45^\circ$ inclination and have four known members. Their semi-major axes are more spread out, $a_{\mbox{\tiny avg}} = 186-297 R_{S}$ , than the Gallic cluster, but is still fairly clustered compared to the large spread of all the Saturnian irregular satellites. The eccentricities are also clustered, with average values $e_{\mbox{\tiny avg}} = 0.29-0.35$. This points to a common origin for the members of this cluster.  

For the Inuit family we observed all the four known members and the broadband colors and spectra are shown in Fig. \ref{fig.colors_inuit} and \ref{fig.spectra_inuit}, respectively. The results show that this cluster is very interesting, showing both dynamical (through the interaction of its members with different resonances) and physical diversity. All the four members of this dynamical family have been reported as possibly caught in resonances \citep{Carruba.2002a,Cuk.2002,Nesvorny.2003}. In order to confirm this we used updated orbits from observations collected by the MPC between 2000-2006 (Marsden, personal communications), and numerically integrated the four objects using a Wisdom-Holman symplectic mapping scheme \citep{Wisdom.1991} for $10^7$ years. The results confirm that S~XXII~Ijiraq and S~XXIV~Kiviuq are found to  reside in the Kozai resonance, while S~XXIX~Siarnaq resides in the secular resonance $g-g_S = 0$, where $g$ and $g_S$ are the satellite and saturnian apsidal frequencies. As reported by \citet{Nesvorny.2003}, we also find that S~XX~Paaliaq does not seem to be associated with the $g-g_S=0$ secular resonance.  

\subsubsection{S~XX~Paaliaq, S~XXIV~Kiviuq and S~XXIX~Siarnaq}
The colors of the three satellites, Siarnaq, Paaliaq and Kiviuq, are very similar with a spectral slope range of ${S'}_2 \sim 10.0 - 13.2$. Paaliaq has a somewhat shallower slope than the two others, but also has the largest uncertainties of the three. The colors are very similar, with matching colors values in $B-V$, $V-R$ and $V-I$ at the $1-$, $2-$ and $3-\sigma$ levels, respectively. All three objects also test positive for the $0.7\mu$m-feature, based on the method proposed by \citet{Vilas.1994a}. The reflectance spectra of these three objects are shown in Fig. \ref{fig.spectra_inuit} and the similarity is easily seen. All the color determinations of these three objects given in this paper is found to have linear slopes at a $2\sigma$ confidence level. 

When comparing the phase curve slope parameter, $G$, of the Inuit cluster it is seen that both Paaliaq and Ijiraq have steeper phase curves than Siarnaq and Kiviuq. The fits for Ijiraq and Kiviuq is poor, however, due to only having two accurate photometric measurements each. The phase curve parameter derived for Siarnaq, $G=0.27\pm0.04$, is lower than reported in \citet{Bauer.2006a}, but the two values are within the $1\sigma$ confidence level of each other. The new value puts Siarnaq in the mid-range for main belt asteroids \citep[$G \sim 0.1-0.5$;][]{Bowell.1989}, but higher than the usual values for C-, P- and D-type asteroids \citep[$G \sim 0.09\pm0.09$;][]{Harris.1989}. The opposition surge of $\sim 0.2$ magnitudes seen in Siarnaq by \citet{Bauer.2006a} is not as narrow in our new fit. In fact none of the irregular satellites observed show any strong opposition surge, consistent with a high albedo surface \citep{Tholen.1994a}.

\subsubsection{S~XXII~Ijiraq}

Ijiraq is distinctly different from the three other members of this dynamical family. Its colors are moderately redder than the others and it does not show any indication of the $0.7\mu$m-feature. The moderately red color found for Ijiraq does not, to our knowledge, appear anywhere in the inner solar system, indicating that its origin is the outer solar system where this color is common. The steep phase curve slope further support the outer solar system as an origin for Ijiraq. With a phase curve parameter of $G=-0.14\pm0.22$ it is very similar to the phase curves of the Centaur population \citep[$G \sim -0.13 - 0.04$;][]{Bauer.2003,Bauer.2006a,Rousselot.2005}. 

\subsection{The Retrograde Satellites}
\label{sec:retrograde}

There are currently 26 known retrograde Saturnian irregular satellites and they show dramatic dynamical and physical variety. The average semi-major axes of these satellites span the range from $213R_S$ to $412R_S$ and they have average inclinations varying from $138^\circ$ to $176^\circ$. Even the average eccentricities show large variety from the low average eccentricity of S~XXIII~Suttungr, $e_{\mbox{\tiny avg}} = 0.115$, to the highly eccentric S/2004~S18,  $e_{\mbox{\tiny avg}} = 0.501$. The retrogrades are thus unlikely be the result of a single capture and subsequent fragmentation. However, it is possible there exist smaller family clusters among the retrograde Saturnian irregular satellites, and using dynamical studies together with the photometric colors and phase curves reported in this paper will make searching for such families possible. The colors of the 5 retrograde irregular satellites that have been observed in this paper are shown in Fig. \ref{fig.colors_nordic}, together with the colors of Phoebe and Ymir \citep{Grav.2003b}. 

\subsubsection{S~XIX~Ymir, S~XXIII~Suttungr, S~XXX~Thrymr}

Comparing the results to earlier observations from \citet{Grav.2003b} reveals some interesting results. Interestingly the B-V color for Ymir has been found to have changed significantly. Furthermore, recent observations by the authors to determine the rotational light curve for Ymir show that the object has a significant peak-to-peak amplitude of $\sim0.3$ magnitudes. It thus seems likely that Ymir has significant surface variegation and most likely an irregular shape. The result from \citet{Grav.2003b} is similar to the colors and slopes of Suttungr and Thrymr found in this paper, both which have similar mean orbital elements to Ymir (see table \ref{taborbele}). This leads us to speculate whether Suttungr and Thrymr might be fragments  broken off Ymir during an impact event. The fact that the two fragments have shallow slope (indicating that their surfaces have not been reddened by exposure to cosmic rays) might be an indication that the impact may have been fairly recent. 

The broadband colors of Suttungr and Thrymr show a sharp dip in the $V$ filter, as can be seen in Fig. \ref{fig.spectra_nordic}. The  While the slope of Thrymr barely can be fitted to a straight line at the $3\sigma$ confidence level giving ${S'}_2 = 1.1\pm0.8$, the dip in the reflectance spectra of Suttungr is significant even at the $3\sigma$ confidence level when using the best fit. A similar, but less significant trend is also seen in the observation of Ymir reported in \citet{Grav.2003b}. We thus believe that the feature is indeed real, but confirmation of this feature is needed either through more accurate broadband photometry or low-resolution spectra. The LRIS B- and V-filters have a central wavelength of $\lambda_c = 436.9$ nm and $\lambda_c =  543.7$ nm with fwhm values of $\Delta\lambda = 88$ nm and $\Delta\lambda = 92$ nm, respectively. It is therefore possible that what is seen is a positive feature at wavelengths just below $400$nm that increases the flux in the B-filter, rather than an absorption feature in the V filter. Such behavior is known among some objects in the cometary population \citep[such as P/Tuttle and IRAS-Araki-Alcock; ][]{Tholen.1984} and the jovian irregular satellites J~XI~Carme \citep{Luu.1991}. Spectra of $H_2O$-, $C_2O$- and $NH_3$-frost all show a significant upturn in the U-filter \citep{Hapke.1981}. All of these types of frost have distinctive features in the near-infrared, thus Ymir, Thrymr and Suttungr are prime candidates for searching for these features. While near-infrared spectra will be nearly impossible to do on these three faint objects, careful preparation might allow for the collection of JHK photometry for these two satellites. Photometric observations in the U-filter would also be very helpful to confirm the possible upturn at low visible wavelengths. 

The notion that these smaller bodies are fragments of Ymir is, however, put into question by a quick look at their orbital elements. Computing dispersion velocities based on Gauss equations we find that Thrymr, S/2006~S4 and S/2004~S8 can be explained with a dispersion velocity of $\delta V \sim 175$ m/s. Suttungr , however, needs a dispersion velocity of $\delta V > 300$ m/s, due to the large difference in eccentricity (from $0.338$ to $0.115$). These dispersion velocities are again significantly higher than what is expected from a fragmentation event \citep{Benz.1999,Pisani.1999,Michel.2002}. Using the formalism from \citet{Nesvorny.2004} and assuming that $\sim 10\%$ of the impact energy is used to disperse the resulting fragments we find that in order for the four fragments to be related to Ymir, the progenitor was impacted by a $\sim 1.75$km large body. The resulting impact energy using the collisional velocity of $4.63$ km/s \citep{Nesvorny.2004} is actually on the same order of magnitude as the energy needed to fully disrupt the progenitor. While the dispersion velocity is high and needs explaining, the significant amplitude of Ymir and the similarities of the colors supports the hypothesis that these objects are indeed the fragments of a catastrophic disruption of a progenitor with a radius of $\sim17$ km.  Again using the \citet{Nesvorny.2004} formalism such a collision has a probability greater then $50\%$ for a protoplanetary disk masses $M_{\mbox{\tiny disk}} \gtrsim 120-160M_\oplus$, depending on the size distribution of the residual planetesimal disk. 

While we have determined phase curve parameters for Ymir, Suttungr and Thrymr in this paper, it should be noted that all fits have problems. Both the fits of Suttungr and Thrymr are very poor, with only one accurate photometric observation each (reported in this paper) and a few measurements reported to the MPC (for which we have assumed an uncertainty of 0.25 magnitudes). Additional photometric observations of these two satellites at a variety of phase angles are necessary in order to better constrain their fits, and thereby allowing us to compare the results to the other irregular satellites. The phase curve fit for Ymir has a low formal error ($G=0.01\pm0.06$), but caution has to be taken. Data collected by the authors, but still in early stages of reduction, show that the rotational light curve has a significant peak-to-peak amplitude. It is therefore surprising to find such a good fit for the phase curve parameters. However, since the determination of any rotational light curve for Ymir is still unconfirmed we stand by the current fit, while cautioning the reader. 

Even with the poor fits it seems clear that Suttungr has a fairly shallow slope ($G=0.68\pm0.46$) compared to the steep slope of Thrymr ($G=-0.22\pm0.39$). This would suggest that the two satellites may have different surface composition and are therefore unlikely to have a common origin.  Furthermore, the phase curve for Ymir is also steep ($G=0.01\pm0.06$). This together with the high dispersion velocity needed to move Suttungr to its present orbit from that of Ymir, we conclude that it unlikely that these two objects have a common origin. It should be noted that the difference in slope between Thrymr and Suttungr has a significance level of only $P=0.136$, making it clear that additional observations and study are essential in order to better understand any relationship between these objects. 

\subsubsection{The neutrally colored objects: S~IX~Phoebe and S~XXV~Mundilfari}

Mundilfari is a small moon, $d \sim 8$km, that has average orbital elements of  $a = 308 R_S$, $e_{\mbox{\tiny avg}} = 0.211$ and $i_{\mbox{\tiny avg}} = 167.0^\circ$. There exists at least two other satellites with similar orbital elements, S/2004~S17 ($320R_S$,$0.181$,$167.8^\circ$) and S/2004~S13 ($303R_S$,$0.259$,$168.8^\circ$) that may have the same origin as Mundilfari. It would be extremely interesting to get color photometry of these two faint satellites, as a similar slope to that of the larger Mundilfari would be strong evidence of a collisional family within the retrograde irregular satellites. 

Mundilfari is found to have a neutral to blueish spectral slope, ${S'}_2 = - 5.0\pm1.9$, which is linear at the $2\sigma$ confidence level. It is seen to have a very shallow feature in the V filter, similar to that of Suttungr and Thrymr. Its colors are similar to the colors of Phoebe at the $2\sigma$ confidence level and the slope is linear at the $2\sigma$ confidence level. The similarity with Phoebe leads us to believe that Mundilfari might have Phoebe as its origin. A dispersion velocity of $\delta V \sim 450$m/s is needed to disperse Mundilfari to its current position, assuming that it is a fragment from a cratering event on Phoebe \citep{Grav.2003b,Nesvorny.2003,Nesvorny.2004}. Again assuming that $\sim10\%$ of the impact energy is used in the dispersion of fragments and a collision velocity of $5.07$ km/s \citep[from ][]{Nesvorny.2004} the impacting body is $2$ km in radius and would excavate a crater with a diameter of $24$km . Images from the Cassini-Huygens space craft show that Phoebe is covered with craters ranging from $100$km in diameter and down \citep{Porco.2005}.  It should be noted that again the dispersion velocity is significantly higher than results from laboratory tests or hydrocode simulations. 

Unfortunately, no other accurate photometric observations exist of Mundilfari. Thus an accurate phase curve can not be determined. Using the inaccurate photometric measurements reported with astrometry to the Minor Planet Center the phase curve is fitted with an essentially flat slope, but the accuracy of the fit makes any comparisons of little worth at this point. Additional observations of Mundilfari at both low and high phase angles are necessary to test whether it has the same phase curve parameter as Phoebe. 

\subsubsection{S~XXVII~Skathi}

This satellite has a spectral slope of ${S'}_2 = 5.2\pm2.8$, which identifies it as a P-type surface. The slope is linear and appears featureless  at the $1\sigma$ uncertainty level. Its colors is very similar to that of Ymir, but it's orbital parameters make any association with Ymir unlikely. It should be noted that Skathi is currently in the region of phase space where secular resonances might have significantly influenced it, but currently its longitude of perihelion, $\varpi$, is circulating in the direction of its orbital motion \citep{Cuk.2004b}. The relationship to the other irregular satellites in its vicinity, particularly S/2004~S19, remains uncertain. Further color photometric observations of the many new retrograde satellites are needed to further explore the existence of possible families, which would help constrain the number of captures needed to create the current distribution. 

\subsection{The lack of ultra red matter among the Saturnian irregular satellites}

It is widely believed that the irregular satellites are captured bodies that were formed in the planetesimal solar disk independent of the planets. The color distribution of the Saturnian irregular satellites give little in suggestions for their source region, but non the less it is generally assumed that the irregular satellites (especially those of the outer planets) have an origin in the outer solar system, as evident by the density of Phoebe \citep{Johnson.2005a}. Figure \ref{fig.spectrahist} shows the spectral distributions of possible source populations to the irregular satellites. The data for the main belt asteroids and trojans was collected using objects in the SDSS Moving Object Catalog with photometry with $\delta m < 0.05$ \citep{Ivezic.2002a}. Values for the outer Solar System populations were computed from observations in \citet{Peixinho.2004a}, \citet{Doressoundiram.2005a} and \citet{Sheppard.2006b}. 

The distributions show that the irregular satellites of Saturn most likely have two sources, one in the main asteroid belt, and one in the outer Solar System. The distribution of the Saturnian irregular satellites is incompatible with both the Centaur, classic Kuiper belt (CKBOs) and Plutino populations, but is in fairly good agreement with the scattered disk population \citep[SDOs; ][]{Doressoundiram.2005a}. The Centaur population also presents a possible source population if one assumes that the it consists of two separate population, one with slopes $0-15\%$ per 100nm and another with slopes $15-30\%$ per 100nm \citep{Peixinho.2003a}. A better understanding of the processes behind the bimodal color distribution among the Centaurs is needed, however, to confidently present its less red population as a source region for the Saturnian irregular satellites. The high number of negative sloped, bluish sloped satellites in our sample is hard to explain from the distribution of Centaurs, SDOs or any other outer Solar system population,  thus leading us to speculate that the inner population of Jovian trojans or main belt asteroids supply a fraction of the Saturnian irregular satellites. This is of course counteracted by the fact that Phoebe is one of these negative, bluish sloped objects. Further study is needed to understand this apparent contradiction. 

However, if we assume that the origin of the Saturnian irregular satellites is in the outer Solar Systems it seems unlikely that no object with a surface covered by ultra red matter (URM) would be captured. No such ultra red matter is seen among the irregular satellites. Ijiraq, with a spectral slope of $S'_2 = 19.5\pm0.9$, is the highest sloped Saturnian irregular satellite. This is significant less then the URM covered surfaces, $S' = 25-50\%$ per 100 nm, found among both the Centaur and scattered disk populations. The same lack of URM is evident in the cometary nuclei, which are also believed to have their origins in the outer solar system \citep{Jewitt.2002a}. Using a collisional resurfacing model with triggered cometary activity \citet{Delsanti.2004a} concluded that most Centaurs should have very red colors, unless their collisional environment is significantly higher than expected. It is likely that the collisional environment of the Saturnian irregular satellites are higher than that of the Centaurs, due to the more constrained phase space they cover and the gravitational focusing of Saturn on heliocentric objects in its vicinity, thus creating less red surfaces due to increased cratering. A more thorough investigation on this issue is beyond the scope of this paper, but is very much needed. 

Since the irregular satellites are believed to have been captured during the later epochs of giant planet formation, it is possible that the lack of URM is simply a result of the time scale of space weathering in the outer Solar System being longer than that of giant planet formation. This means that as objects were scattered inwards from the trans-Neptunian populations and captured by the planets, their surfaces were not yet heavily reddened by space weathering effects. A process is then, however, needed to halt further reddening for objects captured by the giant planets. 

\citet{Jewitt.2002a} hypothesize that the formation of ballistic or rubble mantles by sublimation on outer Solar System objects as they evolve into short period comets is a possible explanation for the lack of URM for the cometary nuclei. Since the Saturnian irregular satellites never enter the water sublimation zone no anticipation of global resurfacing by rubble mantles is expected \citep{Jewitt.2002a}. Ballistic mantles may form on the surfaces of the Saturnian irregular satellites if enough volatiles, in the form of $CO_2$ or other supervolatiles, are exposed near their surfaces. Resurfacing through a ballistic mantel of debris, resulting in a more neutrally colored surface, would then be created by outgassing from active portions of the satellite surfaces. No active outgassing is, however, known among the Saturnian irregular satellites. 

While the low density of Phoebe points to the outer Solar System as the origin, it does not necessarily mean that it accreted in the trans-Neptunian populations. It is quite reasonable that the low density is due to the object originated in the protoplanetary disk between the planets. While the giant planets accreted most of the planetesimals in this region, a significant residual disk could survive for some time after the giant planets have built their cores. It is possible that some of these residuals could be captured into irregular satellites. No currently known objects is believed to have its origin in the region of the giant planets, so it is difficult to determine what the density of such a residual planetesimal would be. It seems, however, unlikely that the density would differ greatly from that of the more distant trans-Nepunian populations. \citet{Sheppard.2006b} speculate that the irregular satellites, Jovian and Neptunian Trojans and the grey excited hot Kuiper belt objects may have a similar origin. Dynamical studies show that these population may be have been populated by objects in the current giant planet domain that were dispersed, transported and trapped in their current locations during or just after the migration of the giant planets \citep{Gomes.2003a,Morbidelli.2005a}.



\subsection{The Norse satellites as the source of Iapetus dark side}
Iapetus is unique among the satellites of Saturn in that the surface on the hemisphere facing in the direction of motion is very dark, on the order of a few percent, while the opposing hemisphere has an albedo roughly ten times greater. Cassini data \citep{Buratti.2005b} have shown that while the NIR spectra of the brightest terrains are consistent with a nearly three-quarters water-ice composition, the darker terrain's 1-5 $\mu$m spectra are consistent with water-ice surface concentrations on the order of a few percent at most. Ferric absorption bands have also been seen in the Cassini spectrum of dark material. Similar bands were reported earlier in ground-based spectra of the dark side of Iapetus, and in spectra of the darkest terrain on Hyperion \citep{Vilas.1996a, Jarvis.2000b}, and were used to show a possible common origin for the dark material on the two satellites. The hypothesis of a similar origin for the dark material of Hyperion and Iapetus was further supported by the successful application of a two-component model (composed of icy and dark D-type asteroidal material) by \citet{Buratti.2002a} in the analysis of high-resolution ground-based optical spectra.  Hence, it has been proposed that the dark material that coats the dark, low-albedo leading side of Iapetus might have an exogenous origin. Furthermore, its origin may be from among the retrograde irregular satellites  \citep[here termed the Norse satellites as they are named after the ice giants in norse mythology;][]{Soter.1974,Cruikshank.1983a, Buratti.1995a,Vilas.1996a,Owen.2001a} and that the same source may possibly coat the surface of Hyperion to a lesser extent. The low-albedo surface of Iapetus has a spectral slope of S$_{2}^{'} \sim$ 7.2\% \citep{Tholen.1983}. Phoebe was the satellite originally proposed as the common source of the dark material, with Poynting-Robertson drag as the potential transport mechanism \citep{Soter.1974}. However, the C-type spectra of this largest irregular satellite \citep{Clark.2005a} is not compatible with the theory that Phoebe is the direct source of the material \citep{Buratti.2005b}.  Some closer matches exist among the irregular satellites from the retrograde group in our color-photometry data. The best candidates for the source of this material thus far are Ymir and Skathi, which exhibit spectral slopes of 5-8 percent, but the future task remains to compare the material spectroscopically.

\section{Conclusions}

\begin{itemize}
\item The Saturnian irregular satellites show a surprising diversity in colors, ranging from neutrally colored Phoebe and Mundilfari, to the moderately red colors of Albiorix and Ijiraq. 
\item None of the observed satellites exhibit ultrared surfaces, as have been observed in the trans-Neptunian populations.
\item The surface of Albiorix appears to be variegated (at a $>3-\sigma$ level). Two of the other Gallic family members we sampled have similar spectral slopes as one portion of Albiorix's surface terrain, and could possibly be products of a collision with the larger irregular satellite. A common origin is furthermore supported by their shared shallow phase curve behavior. 
\item We find a strong agreement in the spectral slope of three of the four Inuit family members we have sampled, possibly indicative of a common origin. 
\item The surface of Ymir also appears to be variegated and may be the parent body of Thrymr and Suttungr, broken apart due to a collision with a $\sim2$ km body.  
\item Our regtrograde sample of satellites show marked diversity. However, two members, Phoebe and Mundilfari, both show uniquely blue slopes, while dynamical arguments support the possibility that they too may have a common origin. 

\end{itemize}

\section{Acknowledgments} 
Based on observations obtained in program GN-2004B-Q-42 and GN-2005A-Q-33 at the Gemini Observatory, which is operated by the Association of Universities for Research in Astronomy, Inc., under a cooperative agreement with the NSF on behalf of the Gemini partnership: the National Science Foundation (United States), the Particle Physics and Astronomy Research Council (United Kingdom), the National Research Council (Canada), CONICYT (Chile), the Australian Research Council (Australia), CNPq (Brazil) and CONICET (Argentina). Some of the data presented herein were obtained at the W. M. Keck Observatory, which is operated as a scientific partnership among the California Institute of Technology, the University of California and the National Aeronautics and Space Administration. The Observatory was made possible by the generous financial support of the W. M. Keck Foundation. This work has been supported by grants NASA/JPL-RSA1264797 and NASA/JPL1270738.

\newpage
\bibliography{../References/references}

\newpage
\section{Table Captions}

\indent\indent{\bf Table \ref{tabphysical}:} The time, instrument and geometry of the observations reported in this paper are shown. The solar, $r$, and observer, $\Delta$, distance of the object is given in AU. The phase angle, $\alpha$, is given in degrees. Notes a,b,c are given in order to better understand which observation of each object corresponds to values given in the subsequent tables.

{\bf Table \ref{tabresults}:} The color photometry with $1\sigma$ uncertainties are given. Notes a,b,c are given in order to better understand which observation of each object corresponds to values given in the other tables.

{\bf Table \ref{tabslopes}:}  The spectral slopes of the observations are given. All slopes are given in $\%$ per $100$nm and all uncertainties are $1\sigma$. Notes a,b,c are given in order to better understand which observation of each object corresponds to values given in the other tables. 

{\bf Table \ref{tabphase}:} Physical parameters computed based on the available observations of the irregular satellites of Saturn. $H(1,1,0)$ and $G$ values of Phoebe taken from \citet{Bauer.2006a}. Mean radius of Phoebe is taken from \citet{Porco.2005}. $0.7\mu$-tests are based on formalism from \cite{Vilas.1994a}, with Y and N reflecting a positive and negative test, respectively.

{\bf Table \ref{taborbele}:} : Based on $10^7$y integrations using a Wisdom-Holman symplectic mapping scheme \citep{Wisdom.1991} based the current best fit orbits from the Minor Planet Center (Marsden, personal communications). 

\FloatBarrier

\begin{table}[p]
\begin{center}
\caption{Observational Geometry}
\label{tabphysical}
\begin{tabular}{lcccccl}
\hline
Object 	& UT 	& r & $\Delta$ & $\alpha$ & Notes\\
\hline
Albiorix	& LRIS-R\&B 2005-Jan-06 10:40 	& 8.9356 & 7.9602 & 0.8545 & a \\
  		& GMOS-N 2005-Apr-13  05:40 	& 8.9742 & 8.9755 & 6.4035 & b \\      
 		& ALFOSC 2001-Feb-18 21:30 	& 9.1750 & 9.2304 & 6.1467 & c, 1) \\
Tarvos  	& GMOS-N 2004-Nov-09 05:10 	& 9.0066 & 8.6380 & 5.9780 & \\
Erriapo 	& GMOS-N 2004-Nov-09 04:00 	& 8.9398 & 8.5467 & 5.9653 &  \\
\hline
Siarnaq 	& LRIS-R\&B 2005-Jan-06 11:47 	& 8.9435 & 7.9686 & 0.8796 & a \\
		& GMOS-N 2005-Apr-15 06:30		& 8.9867 & 9.0185 & 6.3844 & b \\
Paaliaq 	& LRIS-R\&B 2005-Jan-06 10:10 	& 9.0979 & 8.1222 & 0.8277 &   \\
Kiviuq   	& GMOS-N 2004-Nov-10 15:10 	& 9.0183 & 8.6176 & 5.8912 & a \\
		& GMOS-N 2005-Apr-15 06:02		& 9.0199 & 9.0490 & 6.3622 & b \\
Ijiraq     	& GMOS-N 2004-Nov-10 12:51 	& 9.0374 & 8.6361 & 5.8771 & \\
\hline
Ymir      	& LRIS-R\&B 2005-Jan-06 08:10 	& 9.1545 & 8.1811 & 0.9353 & a\\
		& GMOS-N 2005-Apr-13 06:04       	& 9.1871 & 9.1817 & 6.2573 & b \\
Skathi   	& LRIS-R\&B 2005-Jan-06 08:50 	& 9.1157 & 8.1428 & 0.9602 &   \\
Thrymr  	& LRIS-R\&B 2005-Jan-06 09:10 	& 8.8941 & 7.9190 & 0.8776 &  \\
Suttungr 	& LRIS-R\&B 2005-Jan-06 09:36 	& 8.9335 & 7.9585 & 0.8789 &  \\
Mundilfari	& LRIS-R\&B 2005-Jan-06 12:10 	& 9.0969 & 8.1207 & 0.8015 &  \\
\hline
\multicolumn{6}{r}{\small 1) {Re-reduction of observation reported in \citet{Grav.2003b}}} \\

\end{tabular}

\end{center}
\end{table}

\begin{table}[p]
\begin{center}
\caption{Color Photometry}
\label{tabresults}
\begin{tabular}{lccccl}
\hline
Object  & V & B-V & V-R & V-I &  Notes \\
\hline
Albiorix 	& $20.64\pm0.01$ 	& $0.80\pm0.02$ 	& $0.47\pm0.02$ & $0.76\pm0.03$ & a \\ 
                   & $21.32\pm0.01$ 	& $0.96\pm0.02$   	& $0.52\pm0.01$ & $0.96\pm0.02$ & b  \\
           	& $21.40\pm0.07$ 	& $0.79\pm0.12$ 	& $0.52\pm0.07$ & $1.06\pm0.11$ & c,1)\\
Tarvos  	& $23.01\pm0.02$ 	& $0.78\pm0.04$ 	& $0.43\pm0.02$  & $0.82\pm0.03$ \\
Erriapo 	& $23.32\pm0.01$ 	& $0.71\pm0.03$ 	& $0.40\pm0.02$ & $0.86\pm0.02$   \\
\hline
Siarnaq 	& $20.10\pm0.01$ 	& $0.87\pm0.01$ 	& $0.48\pm0.01$ & $1.03\pm0.01$ & a	 \\
                   & $20.72\pm0.01$  	& $0.88\pm0.01$	& $0.49\pm0.01$ & $1.02\pm0.01$ & b   \\
Paaliaq 	& $21.17\pm0.02$ 	& $0.86\pm0.03$ 	& $0.40\pm0.03$ & $0.92\pm0.03$ &	 \\
Kiviuq   	& $22.83\pm0.01$ 	& $0.86\pm0.02$ 	& $0.48\pm0.01$ & $0.98\pm0.02$  & a	 \\
		& $22.63\pm0.02$    & $0.92\pm0.04$     	& $0.50\pm0.03$ & $0.98\pm0.03$  & b \\
Ijiraq     	& $23.40\pm0.01$ 	& $1.05\pm0.03$ 	& $0.58\pm0.02$ & $1.09\pm0.02$ 	& \\
\hline
Ymir      	& $21.88\pm0.02$ 	& $0.80\pm0.03$ 	& $0.45\pm0.02$ & $0.89\pm0.02$ 	& a \\
		& $22.43\pm0.02$	& $0.77\pm0.04$ 	& $0.41\pm0.03$ & $0.86\pm0.04$  & b \\
Skathi   	& $23.80\pm0.05$ 	& $0.72\pm0.08$ 	& $0.37\pm0.06$ & $0.88\pm0.08$  	 \\
Thrymr  	& $23.70\pm0.04$ 	& $0.41\pm0.06$ 	& $0.59\pm0.05$ & $0.86\pm0.07$ 	 \\
Suttungr 	& $24.03\pm0.04$ 	& $0.47\pm0.06$ 	& $0.65\pm0.05$ & $0.78\pm0.06$    \\
Mundilfari & $24.07\pm0.03$ 	& $0.58\pm0.05$ 	& $0.41\pm0.04$ & $0.52\pm0.07$ 	 \\	             
\hline
\multicolumn{6}{r}{\small 1) {Re-reduction of observation reported in \citet{Grav.2003b}}} \\
\end{tabular}

\end{center}
\end{table}

\begin{table}[p]
\begin{center}
\caption{Spectral Slopes}
\label{tabslopes}
\begin{tabular}{lrrrrll}
\hline
Object   & ${S'}_{1}$ & ${S'}_{2}$ & ${S'}_{2(B-R)}$ & ${S'}_{2(R-I)}$ & Taxonomy  & Notes\\
\hline
Albiorix 	& $4.3\pm0.9$ 		& $5.3\pm0.8$		& $11.4\pm1.5$         & $-2.2\pm1.3$          & P & a   \\
		& $12.7\pm0.6$	& $14.9\pm0.5$	& $19.0\pm0.7$         & $5.6\pm1.2$           & D & b\\
		& $11.1\pm3.7$	& $12.9\pm4.1$	& $12.7\pm5.9$         & $12.3\pm9.5$         & D & c, 1)\\
		&				& $12.5\pm0.4$	& $17.3\pm0.6$         & $2.7\pm1.0$        &  & Mean \\
Tarvos  	& $5.3\pm1.2$		& $5.4\pm1.0$		& $ 8.1\pm1.7$          & $2.2\pm2.0$            & P  \\
Erriapo 	& $4.5\pm0.9$		& $5.1\pm0.7$	 	& $ 4.0\pm1.7$          & $6.1\pm1.6$            & P   \\
\hline
Siarnaq 	& $12.2\pm0.3$	& $13.2\pm0.3$ 	& $14.3\pm0.6$ 	& $12.3\pm0.8$ 	& D & a \\
		& $11.8\pm0.8$	& $12.5\pm0.9$	& $9.8\pm1.4$         	& $17.3\pm2.0$        & D & b \\
		&             			& $13.0\pm0.3$	& $13.5\pm0.5$         & $12.6\pm0.7$    &  & Mean \\
Paaliaq     & $9.5\pm1.2$		& $10.0\pm1.2$	& $11.6\pm1.9$	& $9.5\pm3.6$		& D  \\
Kiviuq   	& $10.9\pm0.6$	& $11.4\pm0.6$	& $13.3\pm1.0$ 	& $8.9\pm1.2$		& D& a \\
		& $12.2\pm1.1$	& $12.9\pm1.2$	& $17.1\pm2.0$        & $7.8\pm2.8$           & D & b\\
		&				& $11.8\pm0.6$	& $14.1\pm0.8$        & $8.8\pm1.1$          & & Mean \\
Ijiraq     	& $17.5\pm0.8$ 	& $19.5\pm0.9$	& $25.4\pm1.4$ 	& $11.2\pm2.0$	& MR    \\
\hline
Ymir      	& $7.4\pm0.8$		& $7.4\pm0.7$		& $9.9\pm1.7$		& $5.0\pm2.0$ 		& D & a \\
		& $6.0\pm1.3$		& $6.1\pm1.2$		& $6.9\pm2.1$           & $5.6\pm2.8$           & D & b \\
		&                                   & $8.1\pm0.5$           & $10.0\pm0.9$        & $5.0\pm1.4$        & & Mean \\
Skathi   	 & $5.3\pm2.7$		& $5.2\pm2.8$		& $2.8\pm4.5$ 		& $8.9\pm6.0$		& P  \\
Thrymr  	& $-2.6\pm2.2$		& $-3.0\pm2.7$		&				& 		& C  \\
Suttungr 	 & $-3.1\pm2.0$	& $-3.2\pm2.2$		&				& 		& C   \\
Mundilfari  & $-6.7\pm2.0$	& $-5.0\pm1.9$		& $-1.4\pm3.1$ 	& $-11.7\pm4.0$	 & C  \\
Phoebe	 & $-2.6\pm0.3$ 	& $-2.5\pm0.4$ 	& $-2.5\pm0.7$		& $-2.8\pm0.8$		& C  \\
\hline
\multicolumn{7}{r}{\small 1) {Re-reduction of observation reported in \citet{Grav.2003b}}} \\

\end{tabular}

\end{center}
\end{table}

\begin{table}[p]
\begin{center}
\caption{Physical Parameters}
\label{tabphase}
\begin{tabular}{lrrcc}
\hline
Object  & H(1,1,0) & G & Radius (km) & $0.7\mu$m-test \\
\hline
Albiorix 	& $10.87\pm0.01$ & $0.42\pm0.06$ 	& $\sim16$  	& N $1.060\pm0.005$ \\
Tarvos 	& $12.61\pm0.07$ & $0.19\pm0.15$ 	& $\sim7$ 	& N $1.028\pm0.006$ \\
Erriapo 	& $13.27\pm0.15$ & $0.57\pm0.34$ 	& $\sim5$		& Y $0.974\pm0.011$ \\
\hline
Siarnaq 	& $10.24\pm0.02$ & $0.27\pm0.04$ 	& $\sim21$ 	& Y $0.937\pm0.005$ \\
Paaliaq 	& $11.27\pm0.04$ & $-0.04\pm0.12$ 	& $\sim13$ 	& Y $0.946\pm0.016$ \\
Kiviuq 	& $12.43\pm0.16$ & $0.27\pm0.32$ 	& $\sim8$ 	& Y $0.973\pm0.002$ \\
Ijiraq 	& $12.85\pm0.12$ & $-0.14\pm0.22$     	& $\sim6$ 	& N $1.024\pm0.012$ \\
\hline
Ymir		& $11.81\pm0.02$ & $0.01\pm0.06$ 	& $\sim10$ 	& N $0.997\pm0.011$ \\
Skathi	& $14.04\pm0.11$ & $0.64\pm0.52$		& $\sim4$        & Y $0.910\pm0.016$ \\
Thrymr	& $13.73\pm0.08$ & $-0.22\pm0.30$	& $\sim4$         & N $1.201\pm0.027$ \\
Suttungr	& $14.08\pm0.08$ & $0.68\pm0.46$ 	& $\sim4$         & N $1.337\pm0.041$ \\
Mundilfari & $14.28\pm0.08$ & $0.95\pm0.52$	& $\sim3$         & N $1.146\pm0.146$ \\
Phoebe  	& $  6.24\pm0.01$  & $0.02\pm0.03$        & $106.6\pm1.1$ & N $1.023\pm0.006$\\
\hline
\end{tabular}
\end{center}
\end{table}

\begin{table}[p]

\begin{center}
\caption{Mean Orbital Elements}
\label{taborbele}
\begin{tabular}{lrrr|rrrrrr}
\hline
Object  & $a_{avg}$ & $e_{avg}$ & $i_{avg}$  & $a_{min}$ & $a_{max}$ & $e_{min}$ & $e_{max}$ & $i_{min}$ & $i_{max}$\\
& [AU] & & [$\circ$] & [AU] &  [AU] & & & [$\circ$] & [$\circ$] \\
\hline
\multicolumn{10}{c}{Gallic Satellites} \\
\hline
Albiorix 		& 0.1092 & 0.4768 &  37.16 &  0.1075 & 0.1112 & 0.3256 & 0.6432 & 28.26 & 45.88 \\
Tarvos		& 0.1217 & 0.5254 &  37.93 &  0.1189 & 0.1253 & 0.3580 & 0.7067 & 27.48 & 47.95 \\
Erriapo  		& 0.1170 & 0.4700 &  37.41 &  0.1148 & 0.1199 & 0.3156 & 0.6405 & 28.46 & 45.89 \\
S/2004~S11	& 0.1141 & 0.4684 &  37.88 &  0.1118 & 0.1169 & 0.3075 & 0.6469 & 28.80 & 46.67 \\
\hline
\multicolumn{10}{c}{Inuit Satellites} \\
\hline
Siarnaq		& 0.1195 & 0.2962 &  47.74 &   0.1174 & 0.1220 & 0.0881 & 0.5565 & 40.35 & 54.31\\
Kiviuq		& 0.0756 & 0.3154 &  47.72 &   0.0753 & 0.0759 & 0.1358 & 0.5543 & 39.27 & 53.70 \\
Ijiraq			& 0.0758 & 0.3047 &  48.08 &   0.0755 & 0.0762 & 0.1091 & 0.5659 & 39.12 & 54.03 \\
Paaliaq		& 0.1002 & 0.3457 &  49.24 &   0.0992 & 0.1015 & 0.1101 & 0.6477 & 38.37 & 56.59 \\\hline
\multicolumn{10}{c}{Retrograde Satellites} \\
\hline
S/2004~S18 	& 0.1355 & 0.5006 & 138.42 &  0.1322 & 0.1398 & 0.2576 & 0.7809 & 126.06 & 152.71 \\
Narvi		& 0.1289 & 0.4278 & 142.11 &  0.1266 & 0.1317 & 0.2272 & 0.6673 & 132.40 & 152.76 \\
S/2004~S19	& 0.1226 & 0.3380 & 149.87 &  0.1206 & 0.1251 & 0.1989 & 0.4970 & 142.80 & 156.54 \\
Skathi		& 0.1041 & 0.2737 & 151.93 &  0.1032 & 0.1053 & 0.1792 & 0.3772 & 146.73 & 157.03 \\
S/2006~S2	& 0.1475 & 0.4773 & 153.37 &  0.1432 & 0.1534 & 0.3063 & 0.6599 & 143.48 & 161.98 \\
S/2006~S1	& 0.1251 & 0.1442 & 155.88 &  0.1231 & 0.1273 & 0.0782 & 0.2218 & 151.82 & 159.77 \\
S/2004~S9 	& 0.1354 & 0.2442 & 156.21 &  0.1326 & 0.1390 & 0.1408 & 0.3635 & 150.98 & 161.01 \\
S/2006~S3	& 0.1416 & 0.4511 & 156.75 &  0.1378 & 0.1466 & 0.3018 & 0.6080 & 148.39 & 163.91 \\
S/2004~S15	& 0.1289 & 0.1464 & 158.69 &  0.1267 & 0.1315 & 0.0822 & 0.2225 & 154.72 & 162.51 \\
S/2006~S8	& 0.1178 & 0.4642 & 159.23 &  0.1160 & 0.1203 & 0.3526 & 0.5816 & 152.09 & 165.70 \\
S/2004~S12	& 0.1324 & 0.3289 & 164.62 &  0.1298 & 0.1360 & 0.2322 & 0.4325 & 159.70 & 169.09 \\
S/2004~S16	& 0.1494 & 0.1401 & 164.72 &  0.1456 & 0.1540 & 0.0671 & 0.2240 & 160.90 & 168.39 \\
S/2004~S14	& 0.1316 & 0.3709 & 165.04 &  0.1286 & 0.1356 & 0.2625 & 0.4885 & 159.53 & 169.78 \\
S/2004~S10 	& 0.1388 & 0.2546 & 166.53 &  0.1358 & 0.1427 & 0.1679 & 0.3486 & 162.35 & 170.34 \\
S/2006~S6	& 0.1252 & 0.2175 & 162.79 &  0.1232 & 0.1277 & 0.1477 & 0.2960 & 158.76 & 166.61 \\
Mundilfari		& 0.1243 & 0.2112 & 167.01 &  0.1224 & 0.1265 & 0.1527 & 0.2754 & 163.49 & 170.48 \\
S/2006~S5	& 0.1532 & 0.1897 & 167.66 &  0.1490 & 0.1584 & 0.1049 & 0.2863 & 163.79 & 171.23 \\
S/2004~S17	& 0.1293 & 0.1805 & 167.75 &  0.1270 & 0.1322 & 0.1197 & 0.2483 & 164.25 & 171.15 \\
S/2004~S13	& 0.1226 & 0.2592 & 168.82 &  0.1207 & 0.1251 & 0.1901 & 0.3332 & 165.05 & 172.43 \\
S/2004~S8     	& 0.1661 & 0.2088 & 170.15 &  0.1602 & 0.1735 & 0.1061 & 0.3223 & 166.31 & 173.69 \\
S/2006~S7	& 0.1519 & 0.4470 & 168.77 &  0.1463 & 0.1590 & 0.3074 & 0.5818 & 163.21 & 173.56 \\
Ymir    		& 0.1535 & 0.3376 & 172.86 &  0.1492 & 0.1592 & 0.2311 & 0.4454 & 169.23 & 176.48 \\
S/2006~S4	& 0.1215 & 0.3267 & 174.24 &  0.1196 & 0.1240 & 0.2533 & 0.4013 & 171.09 & 177.53 \\
Phoebe		& 0.0864 & 0.1635 & 174.91 &  0.0860 & 0.0869 & 0.1404 & 0.1883 & 172.52 & 177.92 \\
Suttungr  		& 0.1296 & 0.1152 & 175.69 &  0.1276 & 0.1319 & 0.0734 & 0.1612 & 173.34 & 178.78 \\
Thrymr     		& 0.1360 & 0.4659 & 175.24 &  0.1328 & 0.1402 & 0.3603 & 0.5672 & 171.97 & 178.95 \\
\hline   	

\end{tabular}
\end{center}
\end{table}

\newpage
\section{Figure Captions}

\begin{figure}[p]
\begin{center}
\includegraphics[width=12cm]{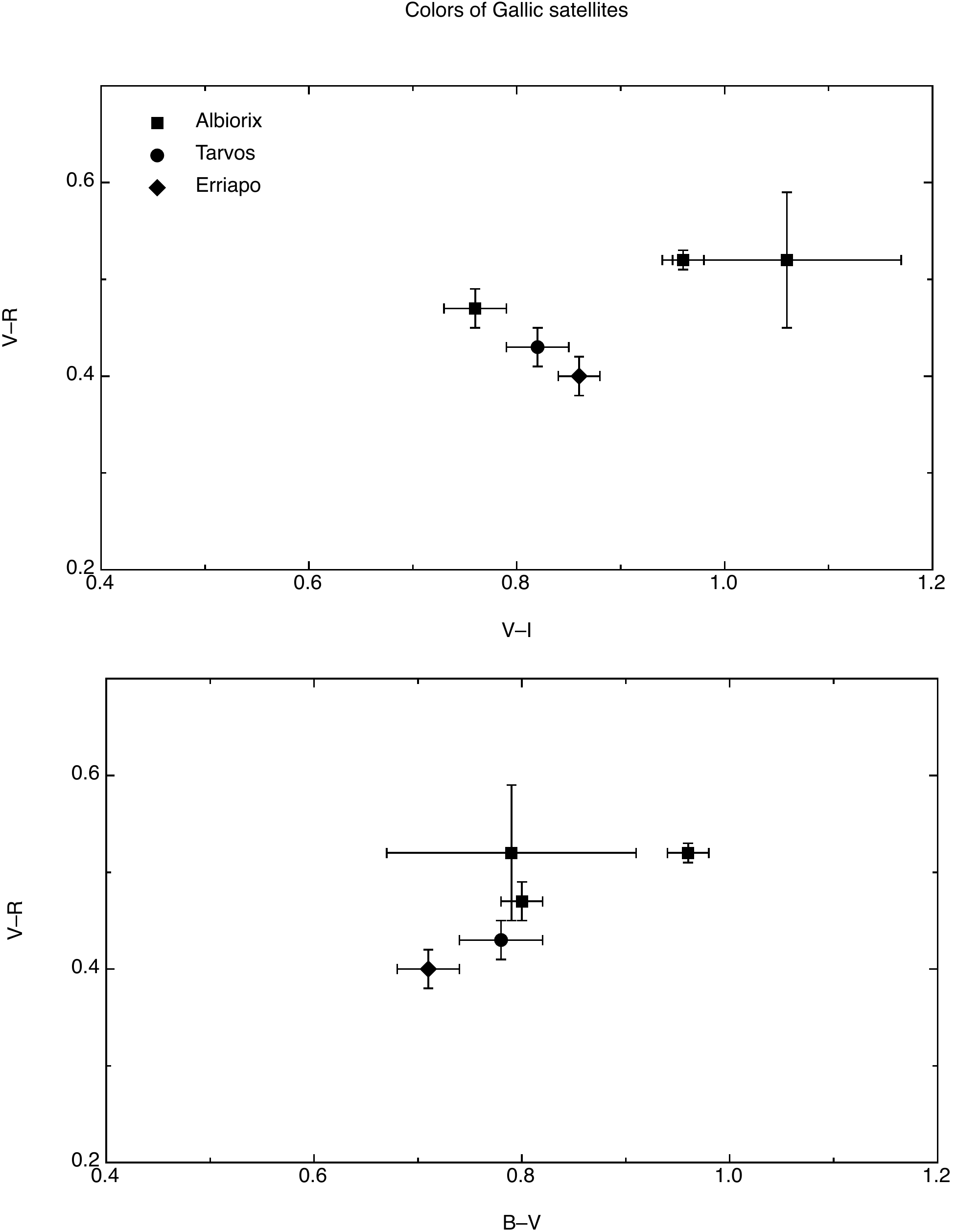}
\caption{}
\label{fig.colors_gallic}
\end{center}
\end{figure}

\indent\indent {\bf Figure \ref{fig.colors_gallic}:} The observed colors for the Gallic family. Both the values from this
work, as well as values from \citet{Buratti.2005} and \citet{Grav.2003b} are plotted. 

\begin{figure}[p]
\begin{center}
\includegraphics[width=12cm]{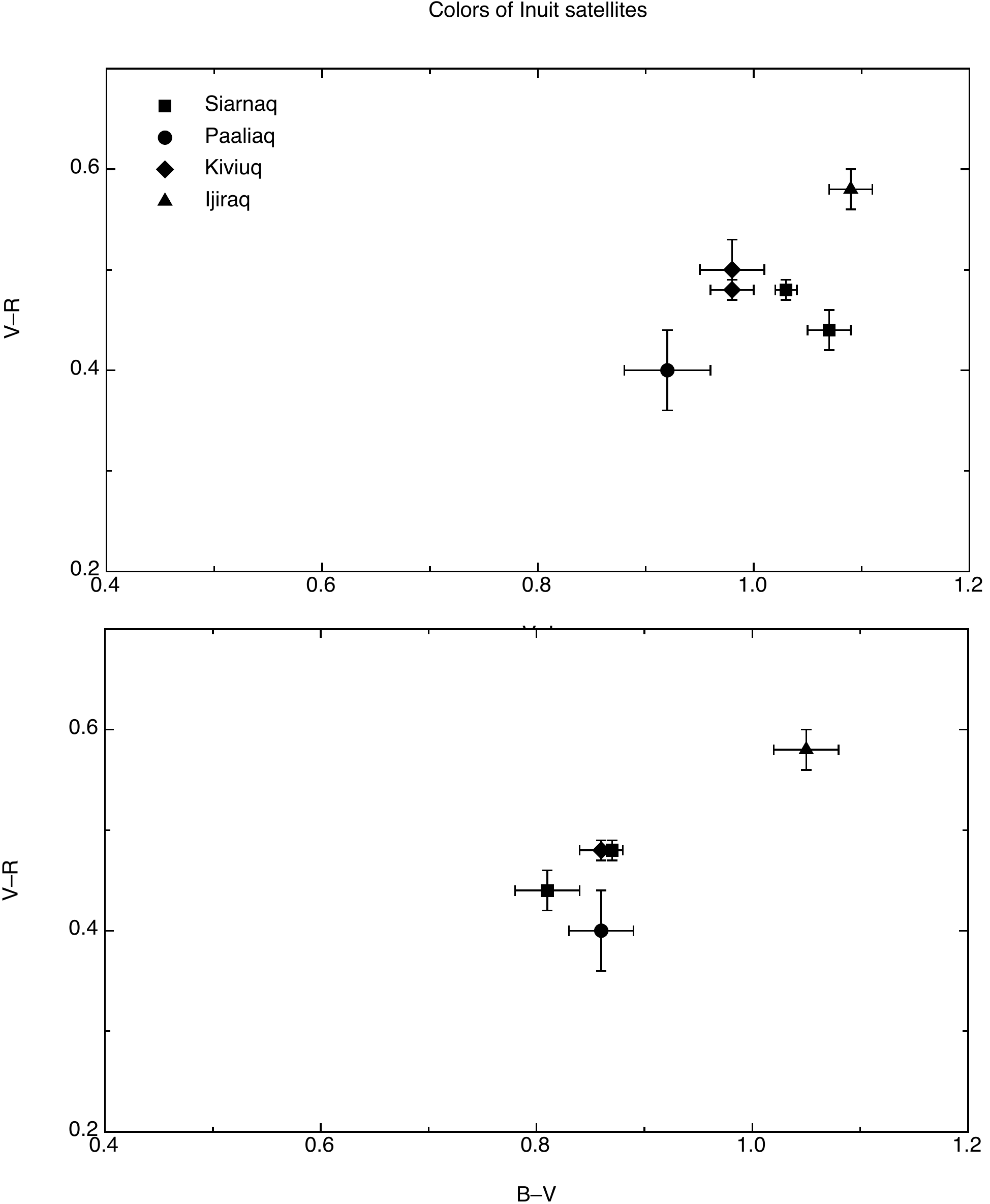}
\caption{}
\label{fig.colors_inuit}
\end{center}
\end{figure}

{\bf Figure \ref{fig.colors_inuit}:} The observed colors for the Inuit family. Both the values from this
work, as well as values from \citet{Buratti.2005} and \citet{Grav.2003b} are plotted.

\begin{figure}[p]
\begin{center}
\includegraphics[width=12cm]{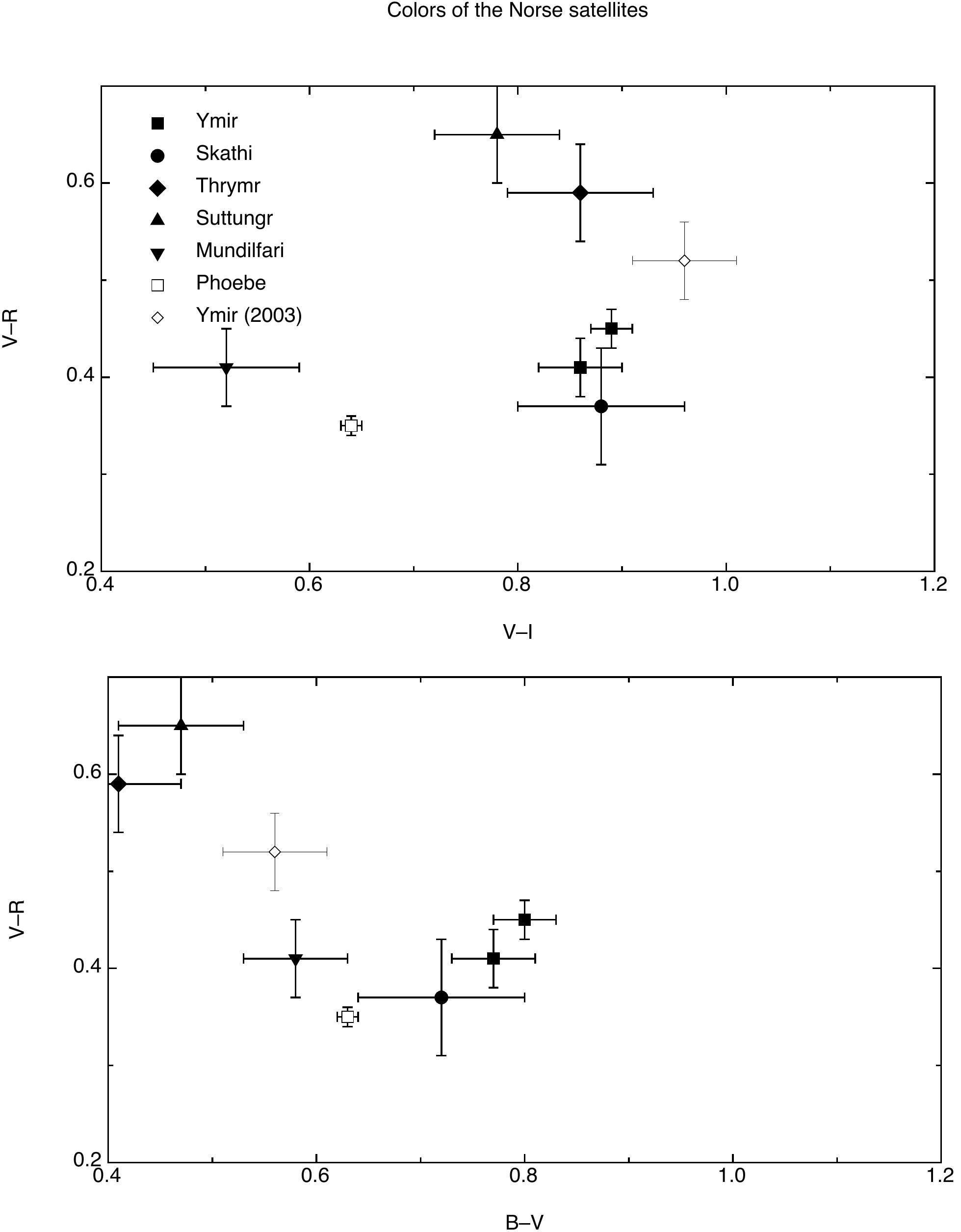}
\caption{}
\label{fig.colors_nordic}
\end{center}
\end{figure}

{\bf Figure \ref{fig.colors_nordic}:} The observed colors for the retrograde Saturnian satellites. Both the values from this
work, as well as values from \citet{Buratti.2005} and \citet{Grav.2003b} are plotted.

\begin{figure}[p]
\begin{center}
\includegraphics[width=12cm]{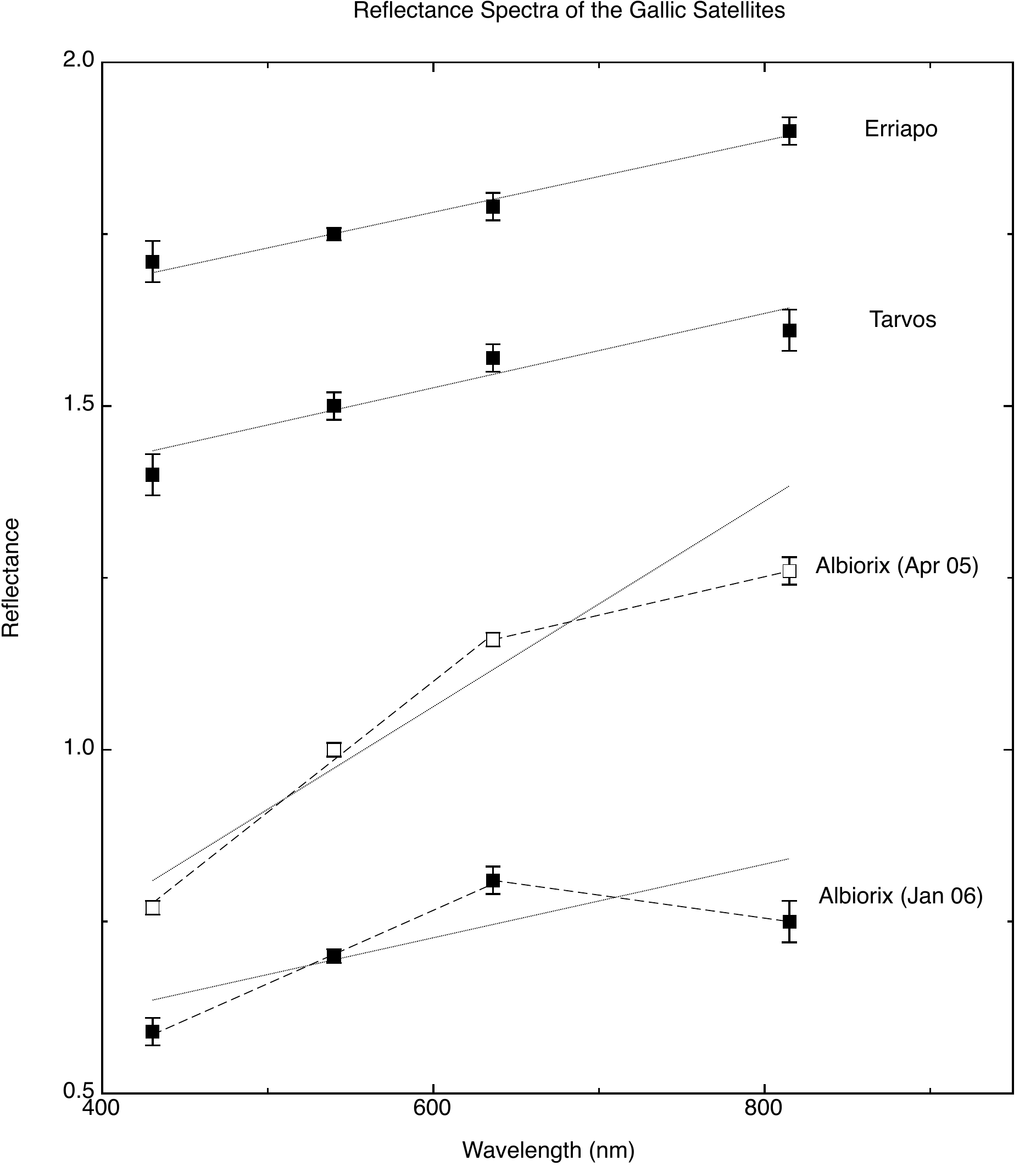}
\caption{}
\label{fig.spectra_gallic}
\end{center}
\end{figure}

{\bf Figure \ref{fig.spectra_gallic}:} The computed reflectance from the observations of the Gallic satellites in this paper are shown. $1\sigma$ uncertainties are shown for all observations and the spectra are offset to better compare their features. The ${S'}_2$ (dotted line), ${S'}_{2(B-R)}$ and ${S'}_{2(R-I)}$ (both dashed lines) are also shown.

\begin{figure}[p]
\begin{center}
\includegraphics[width=12cm]{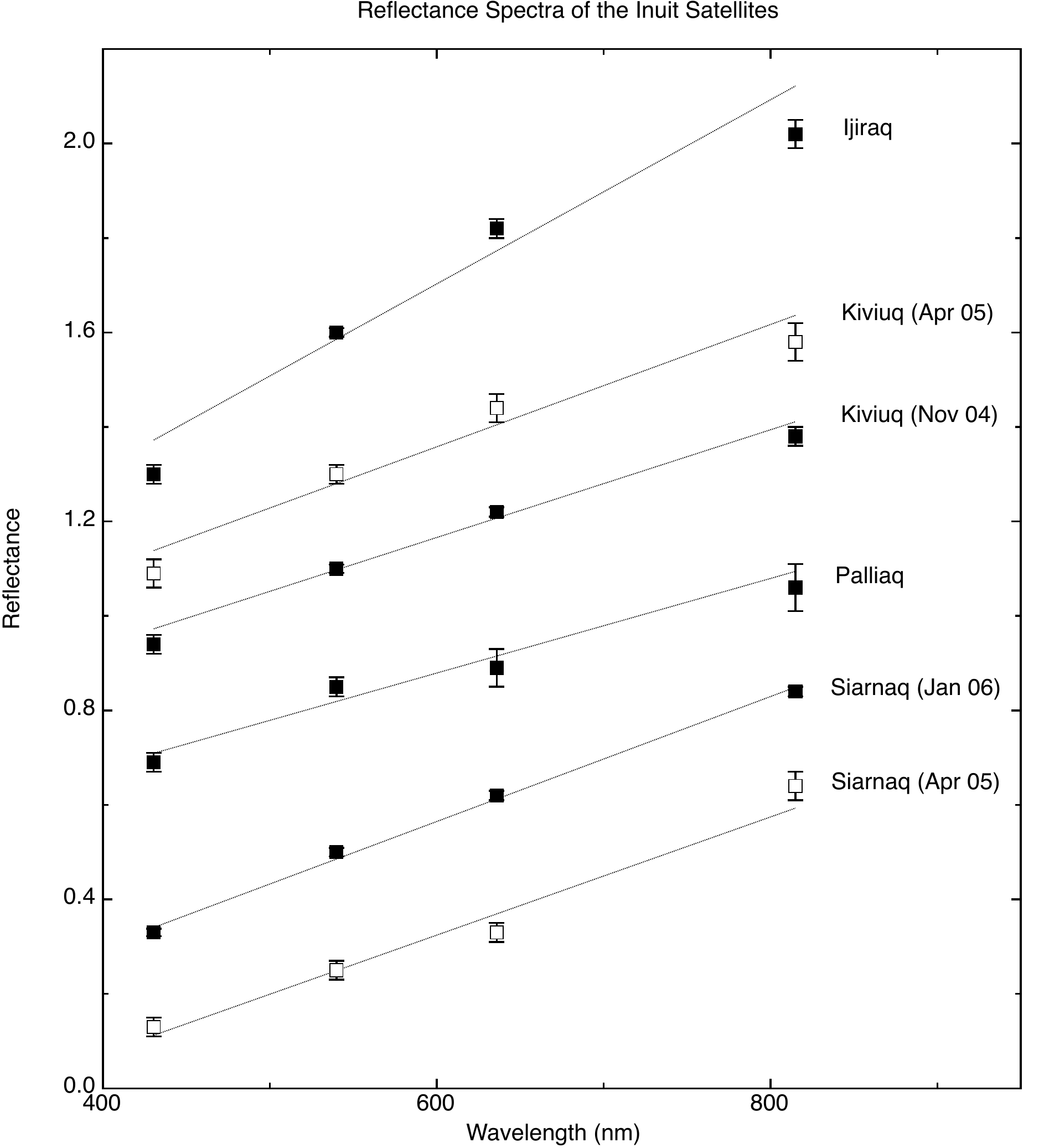}
\caption{}
\label{fig.spectra_inuit}
\end{center}
\end{figure}

{\bf Figure \ref{fig.spectra_inuit}:} The computed reflectance from the observations of the Inuit satellites in this paper are shown. The fitted values for ${S'}_2$ are plotted as dotted lines. $1\sigma$ uncertainties are shown for all observations and the spectra are offset to better compare their features. 

\begin{figure}[p]
\begin{center}
\includegraphics[width=12cm]{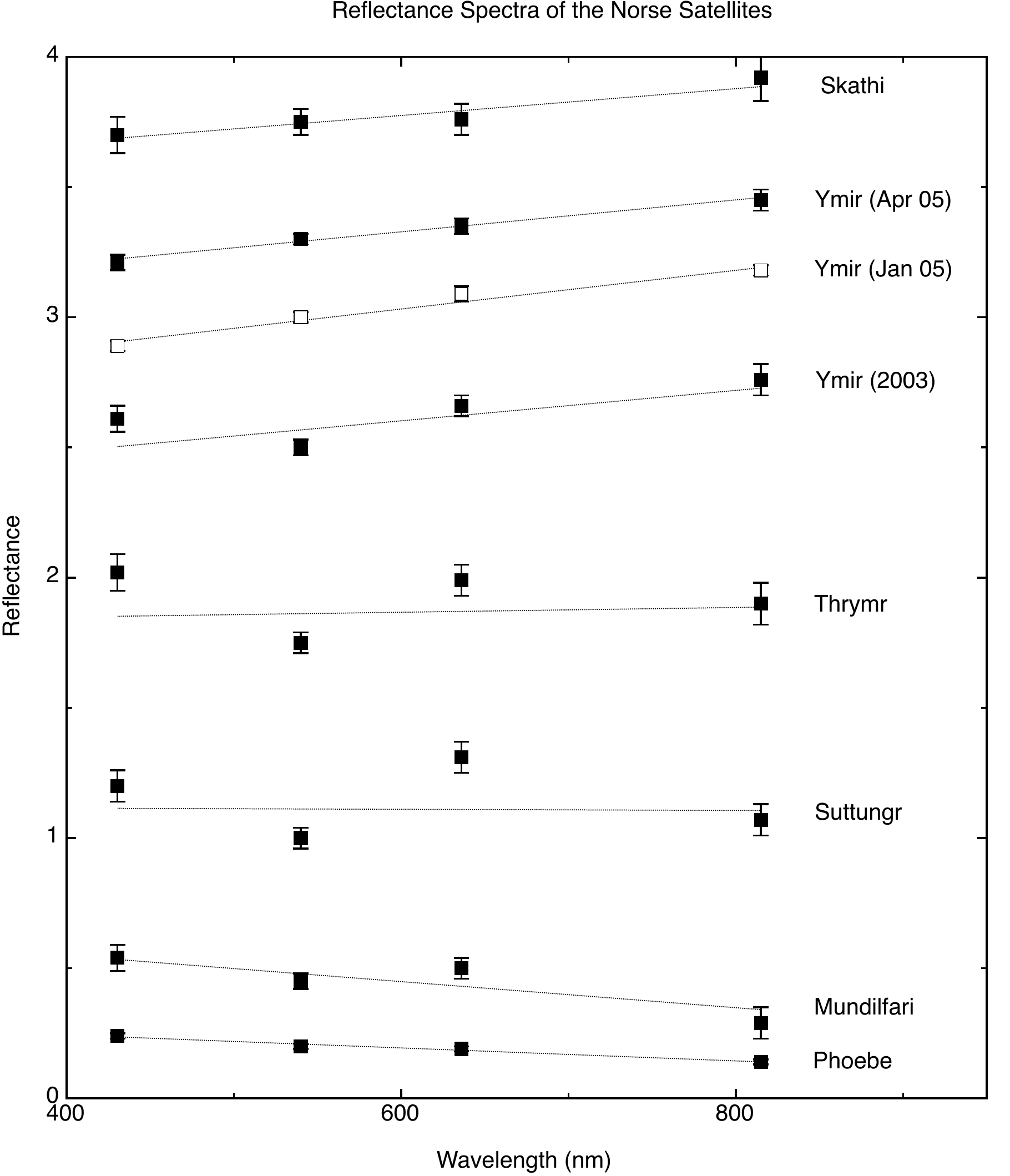}
\caption{}
\label{fig.spectra_nordic}
\end{center}
\end{figure}

{\bf Figure \ref{fig.spectra_nordic}:} The computed reflectance from the observations of the retrograde satellites. The reflectance of Ymir from \citet{Grav.2003b} is also shown. The dotted lines give the fitted $S_2$ spectral slope. $1\sigma$ uncertainties are shown for all observations and the spectra are offset to better compare their features.

\begin{figure}[p]
\begin{center}
\includegraphics[width=12cm]{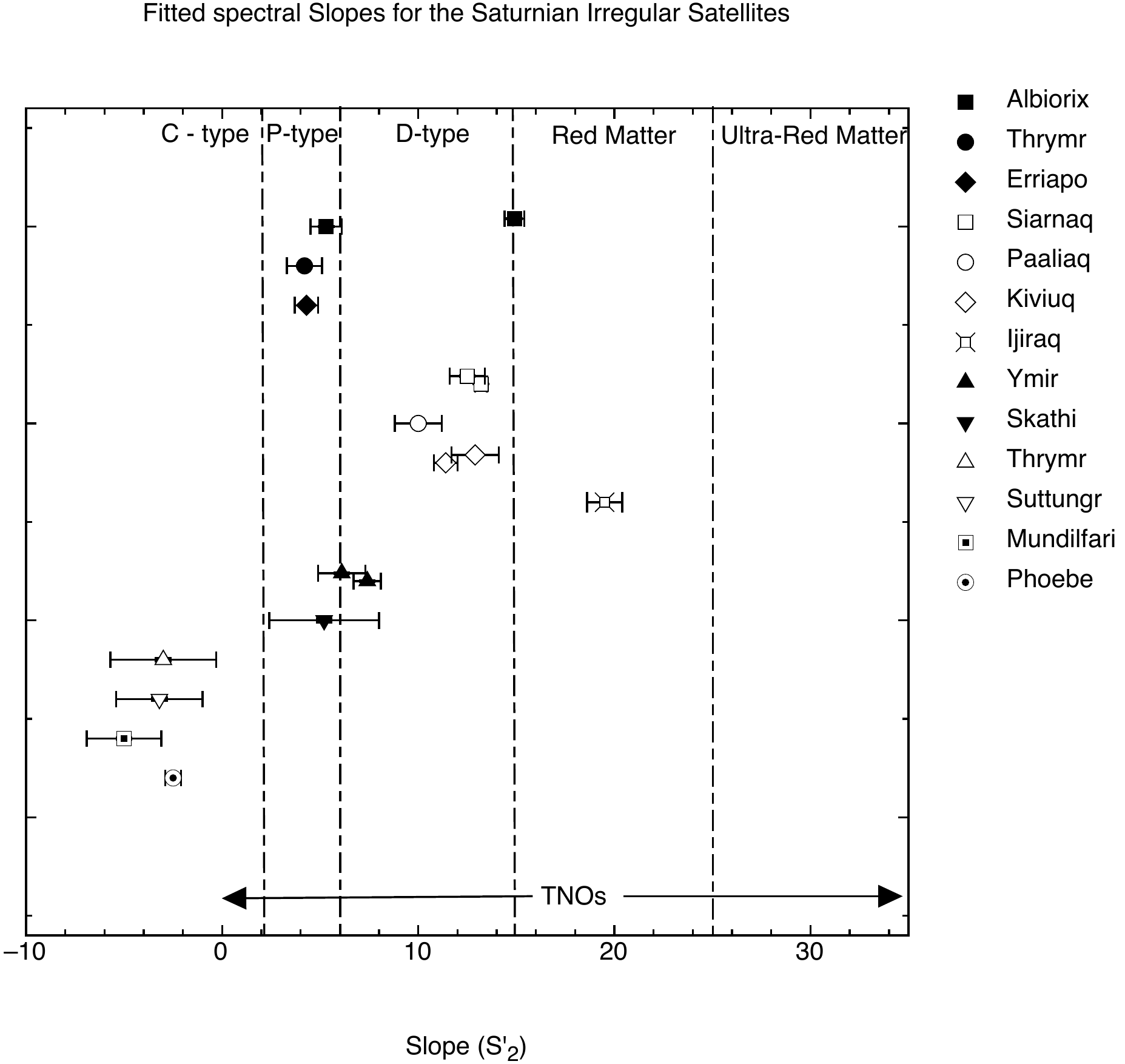}
\caption{}
\label{fig.spectralslope}
\end{center}
\end{figure}

{\bf Figure \ref{fig.spectralslope}:} The fitted spectral slope, ${S'}_2$,  of the Saturnian irregular satellites observed in this paper. Values for Phoebe is calculated from \citet{Grav.2003b}. The slopes of Thrymr and Suttungr are fitted by excluding the V-filter observations. 

\begin{figure}[p]
\begin{center}
\includegraphics[width=12cm]{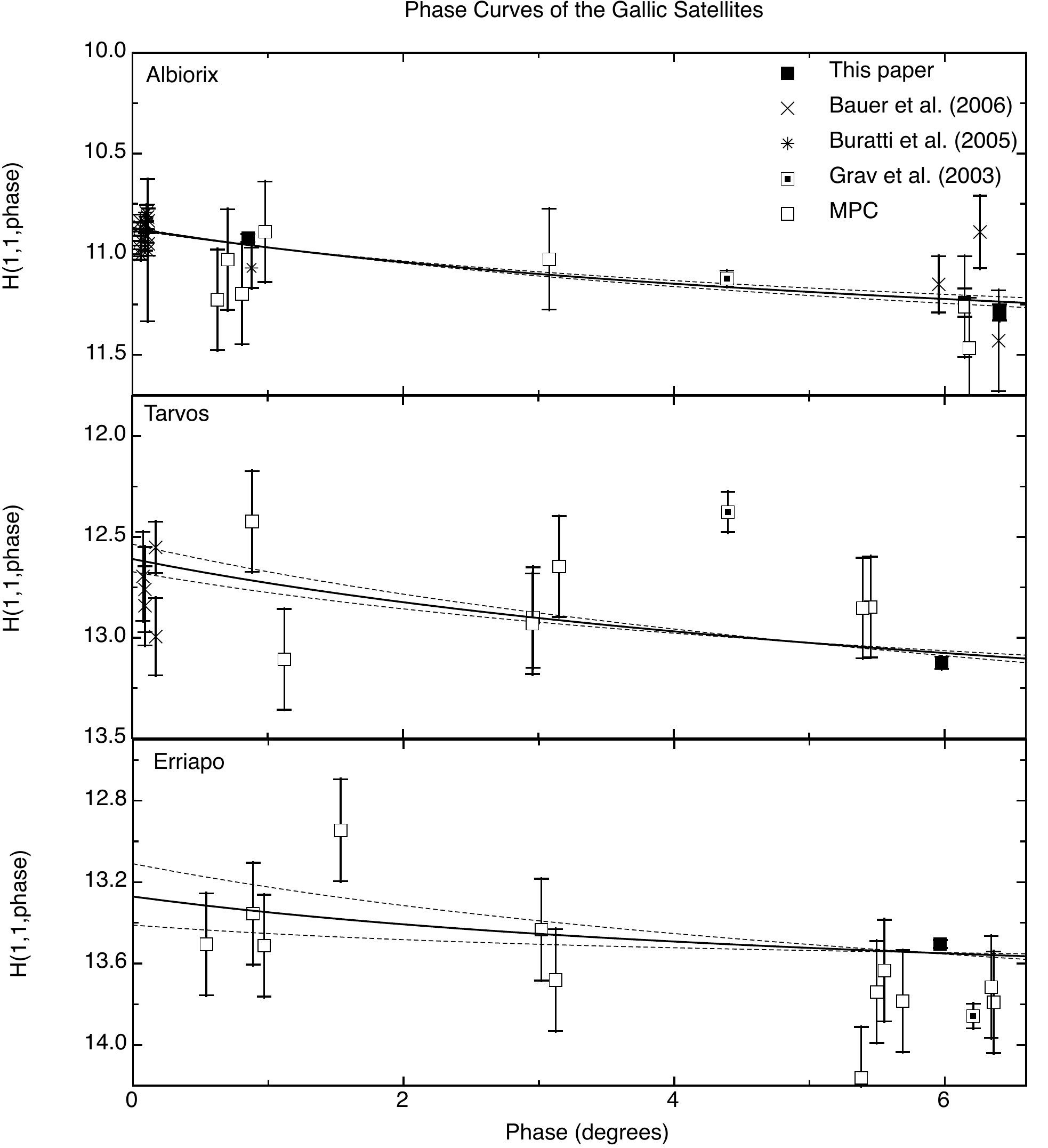}
\caption{}
\label{fig.phasegallic}
\end{center}
\end{figure}

{\bf Figure \ref{fig.phasegallic}:} The available observations are used to determine the IAU phase curve parameters, $H(1,1,0)$ and $G$, for the three Gallic satellites. The best fit (solid line) and error bars (gray lines) are given.

\begin{figure}[p]
\begin{center}
\includegraphics[width=12cm]{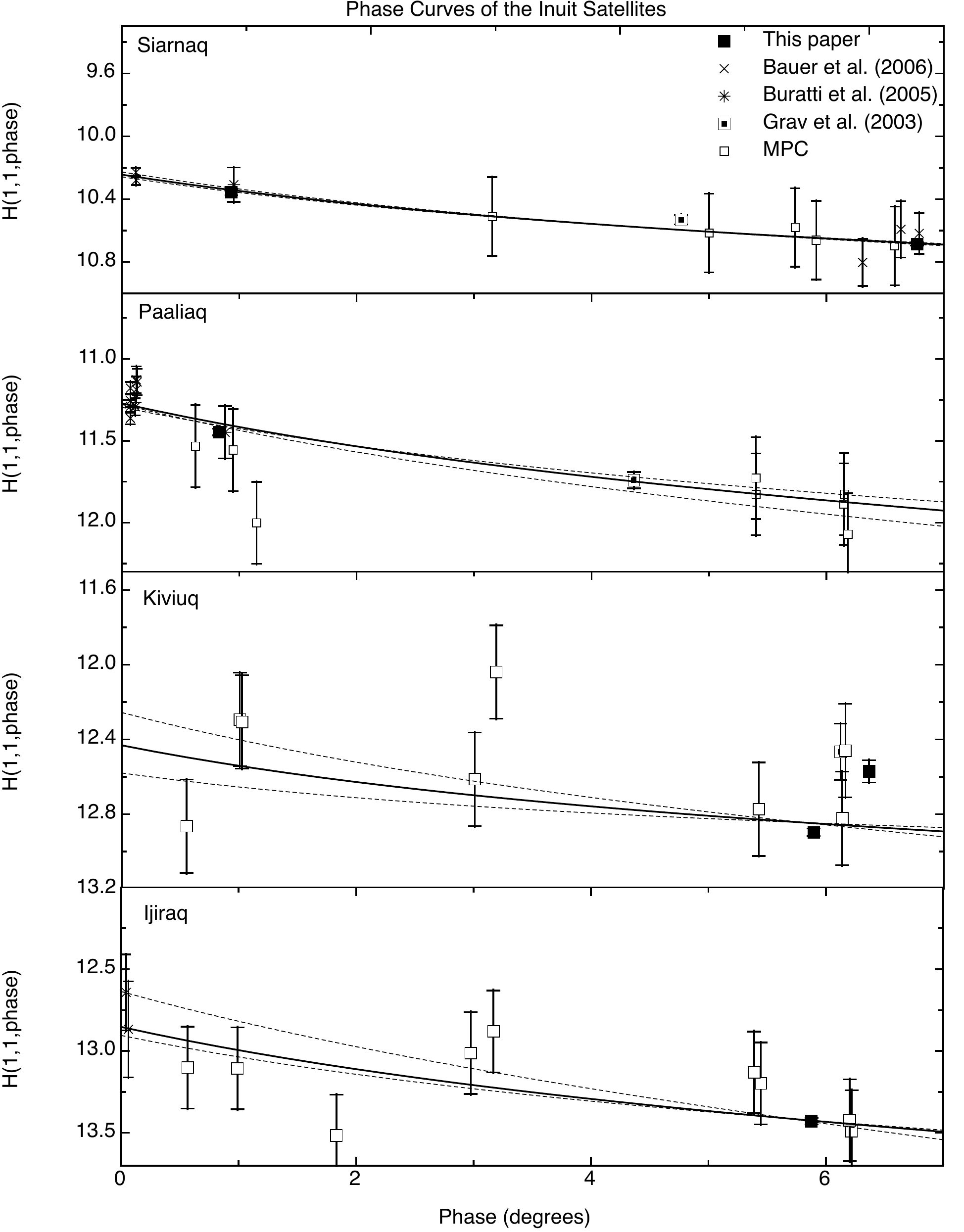}
\caption{}
\label{fig.phaseinuit}
\end{center}
\end{figure}

{\bf Figure \ref{fig.phaseinuit}:} The available observations are used to determine the IAU phase curve parameters, $H(1,1,0)$ and $G$, for the three Inuit satellites. The best fit (solid line) and error bars (gray lines) are given.

\begin{figure}[p]
\begin{center}
\includegraphics[width=12cm]{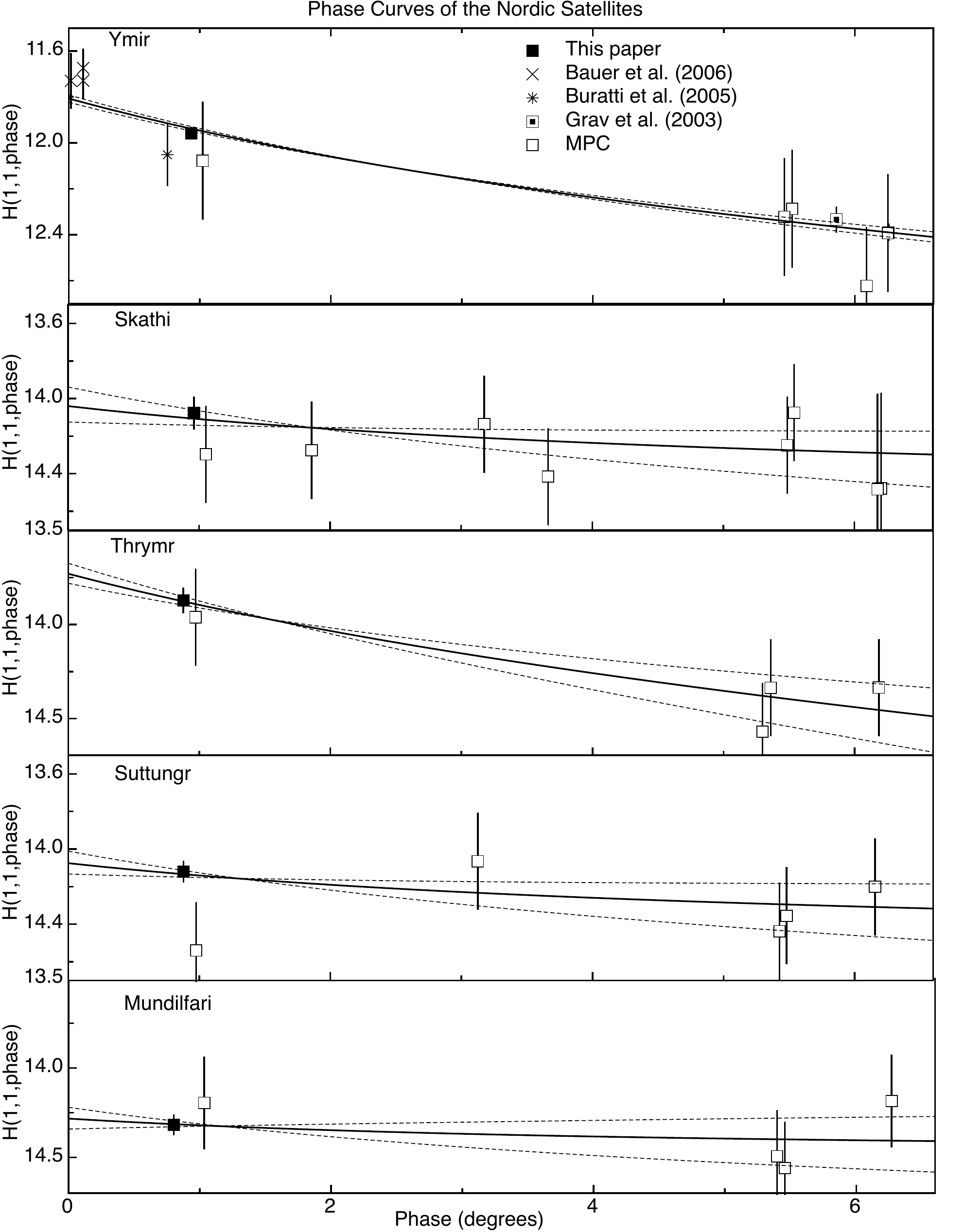}
\caption{}
\label{fig.phasenordic}
\end{center}
\end{figure}

{\bf Figure \ref{fig.phasenordic}:} The available observations are used to determine the IAU phase curve parameters, $H(1,1,0)$ and $G$, for the five Norse satellites. The best fit (solid line) and error bars (gray lines) are given.

\begin{figure}[p]
\begin{center}
\includegraphics[width=12cm]{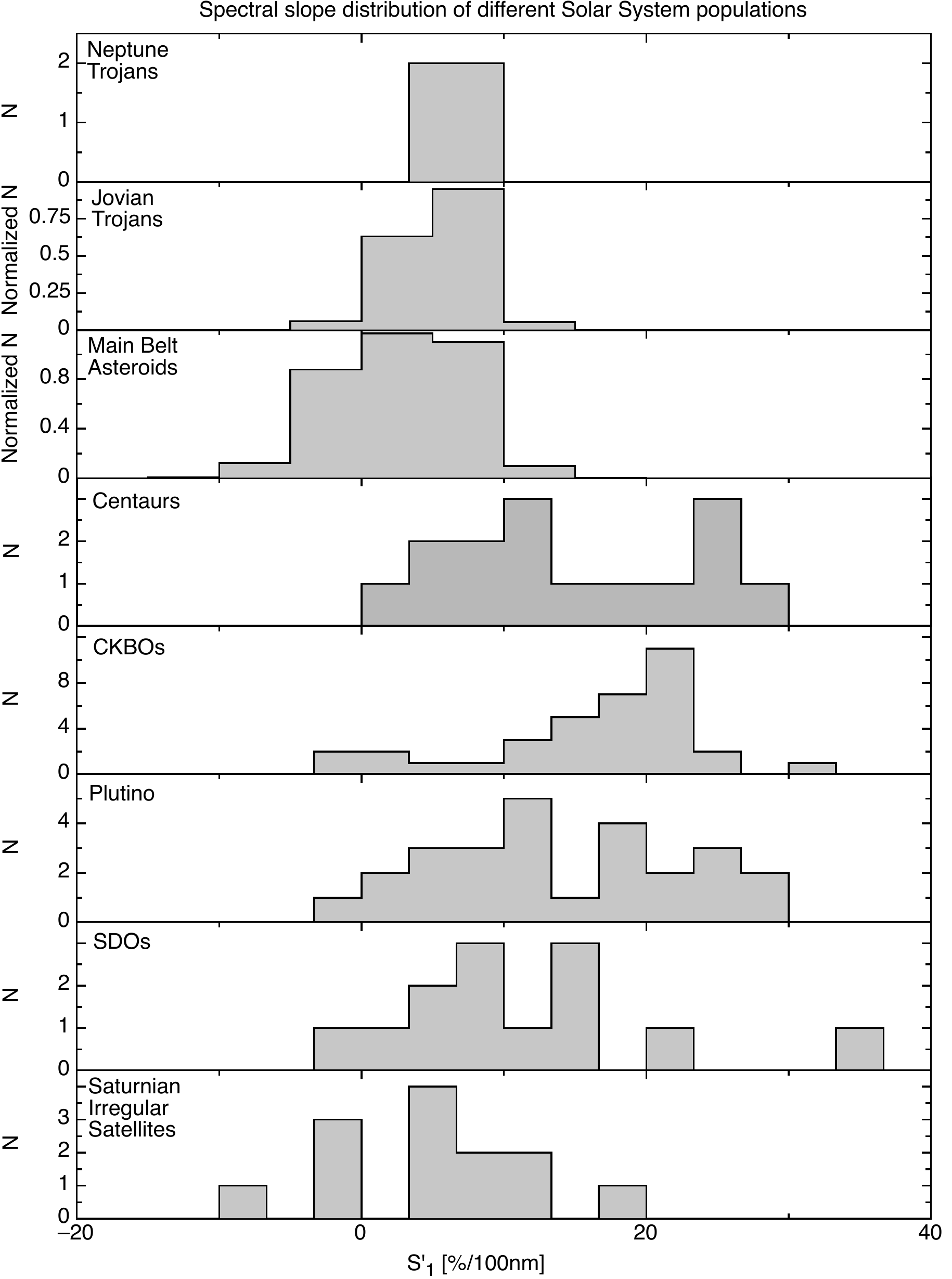}
\caption{}
\label{fig.spectrahist}
\end{center}
\end{figure}

{\bf Figure \ref{fig.spectrahist}:} The spectral slope distributions of several possible origin populations for the Saturnian irregular satellites compared to the distribution found for the irregular satellites in this paper. The ultrared material found in the Centaurs and KBOs does not seem to be present in the Saturnian irregular satellite population. However, the satellites do generally match the bluer component of the Centaurs. The main belt and Jovian Trojan population have been generating using the SDSS Moving Object Catalog \citep{Ivezic.2002a}. The outer Solar System populations have been generated using data from \citet{Peixinho.2004a}, \citet{Doressoundiram.2005a} and \citet{Sheppard.2006b}.

\FloatBarrier

\end{document}